\documentclass[amsmath,amssymb,aps,eqsecnum]{revtex4}
\selectfont 
\usepackage{graphicx} 
\usepackage{subfigure} 
\usepackage{dcolumn} 
\usepackage{bm} 
\usepackage{color}
\usepackage{multirow}
\usepackage{float} 
\usepackage{verbatim}
\usepackage{rotating}

\def\e{\mathrm{e}}

\def\be{\begin{equation}}
\def\ee{\end{equation}}
\def\bea{\begin{eqnarray}}
\def\eea{\end{eqnarray}}

\begin{document}

\title{Scattering of charged fermion to two-dimensional wormhole with constant axial magnetic flux}

\author{Kulapant Pimsamarn}

\email{fsciklpp@ku.ac.th}

\affiliation{Department of Physics, Faculty of Science, Kasetsart University, Bangkok 10900, Thailand}

\author{Piyabut Burikham}

\email{piyabut@gmail.com}

\affiliation{High Energy Physics Theory Group, Department of Physics, Faculty of Science, Chulalongkorn University, Bangkok 10330, Thailand}

\author{Trithos Rojjanason}

\email{trithot@hotmail.com}

\affiliation{High Energy Physics Theory Group, Department of Physics, Faculty of Science, Chulalongkorn University, Bangkok 10330, Thailand}

\begin{abstract}

Scattering of charged fermion with $(1+2)$-dimensional wormhole in the presence of constant axial magnetic flux is explored. By extending the class of fermionic solutions of the Dirac equation in the curved space of wormhole surface to include normal modes with real energy and momentum, we found a quantum selection rule for the scattering of fermion waves to the wormhole. The newly found {\it momentum-angular momentum relation} implies that only fermion with the quantized momentum $k=m'/a\sqrt{q}$ can be transmitted through the hole. The allowed momentum is proportional to an effective angular momentum quantum number $m'$ and inversely proportional to the radius of the throat of the wormhole $a$. Flux dependence of the effective angular momentum quantum number permits us to select fermions that can pass through according to their momenta. A conservation law is also naturally enforced in terms of the unitarity condition among the incident, reflected, and transmitted waves. The scattering involving quasinormal modes~(QNMs) of fermionic states in the wormhole is subsequently explored. It is found that the transmitted waves through the wormhole for all scenarios involving QNMs are mostly suppressed and decaying in time. In the case of QNMs scattering, the unitarity condition is violated but a more generic relation of the scattering coefficients is established. When the magnetic flux $\phi=mhc/e$, i.e., quantized in units of the magnetic flux quantum $hc/e$, the fermion will tunnel through the wormhole with zero reflection.  
 
\end{abstract}

\maketitle

\section{Introduction}\label{sec:intro}

The physics of fermion on 2-dimensional surface has been one of the major research topics in recent years. Interesting boundary phenomena that have no bulk analogs emerge, e.g. quantum~(spin) hall effects, topological matters, and more recently graphene physics~\cite{Novoselov:2005kj}. Two dimensional surface can be manipulated in the laboratory so that it can be curved, strained, and twisted. A cage structure of graphene wormholes, the schwarzite, can even be produced with  promising properties~\cite{schw1,Lherbier}. The gauge fields such as electric~\cite{Novoselov01} and magnetic fields~\cite{Yin} can be applied to the system, leading to unique intriguing 2D phenomena, notably the well-known Landau quantization of fermionic states on a plane. It has been shown that effects of strain and gauge fields could be similar in 2D~\cite{Guinea:2009vd,DeJuan:2013pha}. Behaviour of quantum particle on the curved surface in the presence of gauge fields can be considerably different from the flat situation~\cite{Wang:2014,Wang:2017}. Similar to the strain, curvature effects~\cite{daC1,daC2,BuJe} can mimic gauge fields~\cite{Villarreal,Liang}, specifically curvature and gauge connection appears in the equation of motion with equal role. For example, fermions on 2-dimensional sphere and wormhole experience spin-orbit coupling induced from the surface curvature~\cite{Entin,Rojjanason:2019qpw,Rojjanason,Rojjanason:2018icy} even in the absence of the gauge fields. Addition of axial magnetic field generates Landau quantization distinctive from the planar case. For the wormhole, the fermionic states can be in quasinormal modes with complex energies, its quantum statistics can be altered by the magnetic flux through the hole~\cite{Wilczek,Rojjanason:2018icy}. 
   
Dirac fermion in graphene wormhole without gauge field has been discussed in Ref.~\cite{Gonzalez,IoLa}. In Ref.~\cite{Rojjanason:2018icy}, we investigate the effects of axial magnetic field on a charged fermion in a $(1+2)$-dimensional wormhole.  This 2-dimensional wormhole is fundamentally different from the $(1+3)$-dimensional wormhole in General Relativity~(see e.g. \cite{Konoplya2005}), there is no time dilation in the 2-dimensional wormhole under consideration. For the constant magnetic flux scenario, the system can be solved analytically and exact solutions are found to contain ``normal''~(real energy but complex momentum) and quasinormal modes~(QNMs).  

In this work the scattering of fermions to the wormhole is explored. The more generic solutions to the equation of motion of the fermion in the magnetized wormhole are constructed in terms of hypergeometric functions. The normal modes are actually found when the (effective) angular quantum number $m'$~(see definition below) is related to the momentum of the fermion by $k = \displaystyle{\frac{m'}{a\sqrt{q}}}\equiv k_{m'}$. Such quantized {\it momentum-angular momentum relation} is unique to the 2D wormhole under consideration. 

In Section \ref{sec:Line element}, the mathematical formulation of the wormhole and fermion in curved space is established. In Section \ref{sec:DiracEq} as a review of the main results of Ref.~\cite{Rojjanason:2018icy}, the Dirac equation in the magnetized wormhole is written and solved analytically, then the general solutions in the upper and lower plane connected to the wormhole is discussed. Matching conditions of the scattering of fermionic waves to the wormhole is considered in Section \ref{secOutWormhole}. Section \ref{secgenrel} discusses the use of Wronskian to derive a general relation between scattering coefficients. The scattering scenarios are categorized into normal-modes and QNMs scattering and some of the results are presented in Section \ref{secScat}. Section \ref{secCon} concludes our work.

\section{\label{sec:Line element} Geometric and gauge setup of the wormhole}

\begin{figure}[H]
  \centering
  \includegraphics[width=0.6\textwidth]{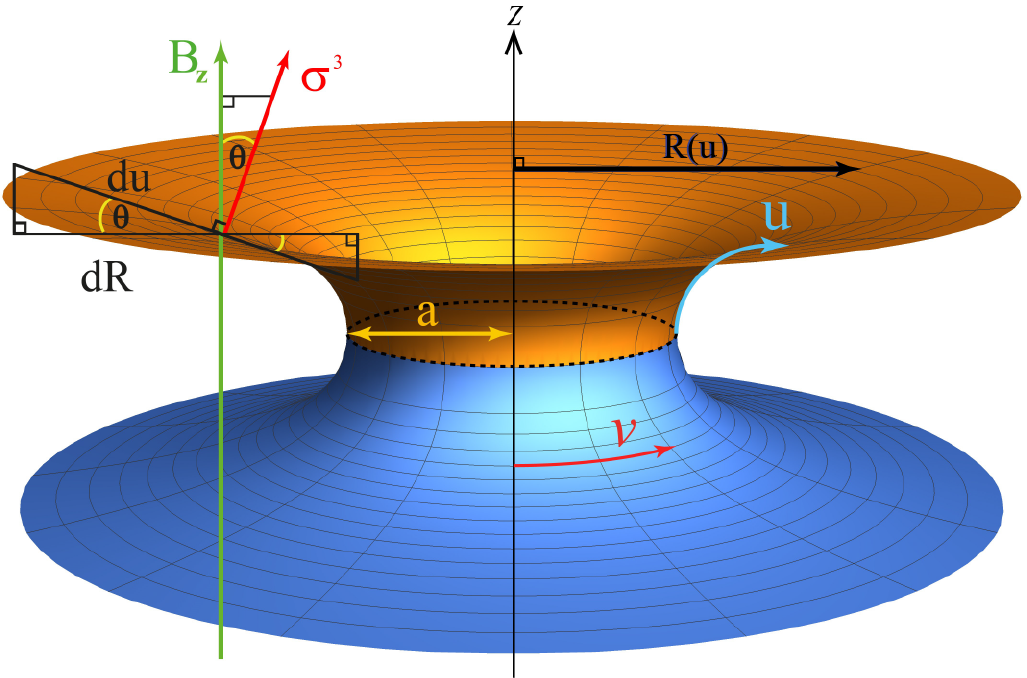}
  \caption{Geometric structure of a wormhole surface where $a$ is a radius at the radius function $R(u=u_{0}=\ln q/2)$. And $r$ is the radius of curvature of the wormhole surface along $u$ direction. }  \label{fig1}
\end{figure}

In the (1+2)-dimensional curved spacetime, the line element on the surface of wormhole can be cast in the following form
\begin{equation}\label{eq:7}
ds^{2}= g_{\mu \upsilon} dx^{\mu} dx^{\upsilon}=-c^{2}dt^{2}+du^{2}+R^{2}(u)d v^{2}.
\end{equation}
The fermion will experience the effective curvature that can be addressed by considering the Dirac equation in curved spacetime
\begin{equation}\label{eq:EOM01}
\left[\gamma^{a}e^{\mu}_{a}\left(-\hbar\nabla_{\mu}+i\frac{e}{c}A_{\mu}\right)-Mc\right]\Psi = 0,
\end{equation}
where $\Psi=\Psi(t,u,v)$ represents the Dirac spinor field on the wormhole and $M$ represents the rest mass of the particle, $c$ is the speed of light, $e$ is electric charge, and $A_{\mu}$ is the electromagnetic four-potential.  The $\gamma^{a}$ are the Dirac matrices given by
\begin{equation}\nonumber
\gamma^{0}=\left(
 \begin{array}{cc}
 i & 0 \\
  0 & -i \\
  \end{array}
 \right)
,\quad
\gamma^{k}=\left(
  \begin{array}{cc}
  0 & i\sigma^{k} \\
  -i\sigma^{k} & 0 \\
  \end{array}
 \right),
\end{equation}
where $\sigma^{k}$ are the Pauli matrices defined by
\begin{equation}\nonumber
 \sigma^{1}=\left(
 \begin{array}{cc}
 0 & 1 \\
  1 & 0 \\
  \end{array}
 \right)
,\quad
\sigma^{2}=\left(
  \begin{array}{cc}
  0 & -i \\
  i & 0 \\
  \end{array}
 \right),\quad
 \sigma^{3}=\left(
  \begin{array}{cc}
  1 & 0 \\
  0 & -1 \\
  \end{array}
\right).
\end{equation}
They obey the Clifford algebra
\begin{equation}\label{eq:10}
\{\gamma^{a}, \gamma^{b}\}=2\eta^{ab}.
\end{equation}
The Pauli matrices have a useful identity that we will use later
\begin{equation}\label{eq:11}
\sigma^{i}\sigma^{j}=\delta^{ij}+i\epsilon^{ijk}\sigma^{k},
\end{equation}
where $\epsilon^{ijk}$ is Levi-Civita symbol.

The covariant derivative of the spinor interaction with gauge field in the curved space is given by
\begin{equation} \label{eq:12}
\nabla_{\mu}\equiv \partial_{\mu}-\Gamma_{\mu},
\end{equation}
where the spin connection $\Gamma_{\mu}$ \cite{Yepez} is
\begin{equation}\label{eq:11}
\Gamma_{\mu}=-\frac{1}{4} \gamma^{a}\gamma^{b} e^{\nu}_{a} \left[\partial_{\mu}\left(g_{\nu \beta}e^{\beta}_{b}\right)-e^{\beta}_{b}\Gamma_{\beta\mu\nu} \right],
\end{equation}
where $\beta,\mu,\nu \in \{t,u,v \}$ and the Christoffel symbols $\Gamma_{\beta \mu \nu}$ are defined by
\begin{equation}\nonumber
\Gamma_{\beta \mu \nu}=\frac{1}{2}\left(\partial_{\mu}g_{\beta \nu}+\partial_{\nu}g_{\beta \mu}-\partial_{\beta}g_{\mu \nu}\right).
\end{equation}
Then for the metric (\ref{eq:7}),
\begin{equation}\label{eq:12}
-\Gamma_{uvv} = \Gamma_{vuv} = \Gamma_{vvu}=\frac{1}{2}\partial_{u}R^{2}=RR',
\end{equation} 
and  zero otherwise. It is then straightforward to show that the spin connections are
\begin{equation}\label{eq:13}
\Gamma_{t}=0, \quad
\Gamma_{u}=0, \quad
\Gamma_{v}=\frac{1}{2}\gamma^{1}\gamma^{2}R'.
\end{equation}
We will apply an external magnetic field such that the $z$-component $B_{z}=B(z)$ is uniform with respect to the plane $(x,y)$ and the magnetic {\it flux} through the circular area enclosed by the wormhole at a fixed $z$ is constant, namely $B_{z}\sim 1/R^{2}$. Due to the axial symmetry, the electromagnetic four-potential can be expressed in the axial gauge as
\begin{equation}\nonumber
A_{\mu'}(t,x,y,z)=(0,-\frac{1}{2}By,\frac{1}{2}Bx,0),
\end{equation}
and in the wormhole coordinates as
\begin{equation}\label{eq:Gauge2}
A_{\mu}(t,u,v)=\frac{\partial_{}x^{\nu'}}{\partial_{}x^{\mu}}A_{\nu'}(t,x,y,z)=\left(0,0,\frac{1}{2}BR^{2}\right).
\end{equation}
The magnetic field is then given by
\be
\vec{B} = \left(-\frac{x}{2}\partial_{z} B, -\frac{y}{2}\partial_{z} B, B\right).
\ee
We will now consider the Dirac equation of fermion in the wormhole in the presence of constant magnetic flux.

\section{\label{sec:DiracEq}The Dirac equation in magnetized wormhole}

Utilizing the results from above equations, the Dirac equation Eq.(\ref{eq:EOM01}) can be written in the form then
\be
\left(\begin{array}{cc}
 (i\partial_{ct}+\frac{Mc}{\hbar}) & i\textbf{D} \\
  -i\textbf{D} & (-i\partial_{ct}+\frac{Mc}{\hbar}) \\
  \end{array}\right)\Psi=0,   \label{eq:EOM02}
\end{equation} 
where $\textbf{D}$ is a differential operator 
\begin{equation}\label{eq:21}
\textbf{D}\equiv \sigma^{1}\left(\partial_{u}+ \frac{R'}{2R}\right)+\sigma^{2}\left(\frac{1}{R}\partial_{v}-\frac{ie}{2\hbar c}BR\right).
\end{equation}
We can define the pseudo-vector potential as
\begin{equation}
\textbf{D}=\sigma^{1}\Big(\partial_{u}-i\frac{e}{\hbar c}A_{\tilde{u}}\Big)+ \sigma^{2}\frac{1}{R}\Big(\partial_{v}-i\frac{e}{\hbar c}A_{v}\Big)\quad    \\
;\quad A_{\tilde{u}}\equiv i\frac{\hbar c}{e}\frac{R'}{2R},\quad A_{u}=0,\quad A_{v}=\frac{1}{2}BR^{2}.  \label{coneq}
\ee
Note that the effective gauge potential in the $u$ direction, $A_{\tilde{u}}$, is generated by the curvature along the $v$ direction, $\Gamma_{v}$.  In this sense, wormhole ``gravity'' or curvature connection manifests itself in the form of gauge connection in the perpendicular direction.

Consider a stationary state of the Dirac spinor needs to be single-valued at every point in spacetime, $\Psi(t,u,v)$ must be a periodic function in $v$ with period $v\in[0,2\pi]$, in the form
\begin{equation}\label{eq:Solution}
\Psi(t,u,v)=\e^{-\frac{i}{\hbar}Et}\e^{imv}\left(
\begin{array}{cc}
  \chi(u) \\
  \varphi(u)
  \end{array}
\right),
\end{equation}
where the orbital angular momentum quantum number $m=0,\pm 1,\pm 2,...$. $\chi(u),\varphi(u)$ are two-component spinors.  Eq.(\ref{eq:EOM02}) can be rewritten in the form of coupled equations for the 2-spinors
\begin{equation}\label{eq:19}
\left(\frac{E+Mc^{2}}{\hbar c}\right)\chi(u)+i\textbf{D}\varphi(u)=0
\end{equation}
\begin{equation}\label{eq:20}
\left(\frac{E-Mc^{2}}{\hbar c}\right)\varphi(u)+i\textbf{D}\chi(u) =0
\end{equation}

From Eq.(\ref{eq:21}), the first term is equivalent to the Dirac operator with the pseudo gauge potential $A_{\tilde{u}}(u)$ in the $u$ direction. This term is generated from the change in the radius of the hole. The second term contains gauge potential that generates a spin-orbit coupling. A similar setup has been used to study nanotubes under a sinusoidal potential \cite{Gonzalez}. Here we consider the dispersion relation for the two-dimensional fermions described by the Dirac equation in the presence of the effective potential arising from the wormhole geometrical structure. In the presence of external magnetic field $\mathbf{B}=\bigtriangledown\times\mathbf{A}$ along the $z$ direction, the charged fermion moving in $v$ direction is expected to form a stationary state with quantized angular momentum and energy, i.e. the Landau levels in the curved space with hole.  To show this, we need to solve for the stationary states of the system.  We can start by considering $-i\textbf{D}\times$Eq.(\ref{eq:19})-$\left[\left(E+Mc^{2}\right)/\hbar c\right] \times$Eq.(\ref{eq:20}) to obtain
\begin{equation}\label{eq:EOMDirac01}
\begin{split}
\left[\textbf{D}^{2}+\frac{E^2-M^2c^4}{\hbar^2 c^2}\right]\varphi(u)
              &=0.
\end{split}
\end{equation}
For constant magnetic flux, the operator $\textbf{D}^{2}$ in the equation of motion now takes the form
\begin{equation}\label{Dsq}
\begin{split}
\textbf{D}^2
          &=\partial_{u}^2
          +\frac{R'}{R}\partial_{u}
          -\Big(\frac{R'}{2R}\Big)^2
          +\frac{R''}{2R}
          +\frac{1}{R^2}\Big(\partial_{v}-i\frac{\phi}{\phi_0}\Big)^2
          -i\sigma^{3}\frac{R'}{R^2}\Big(\partial_{v}-i\frac{\phi}{\phi_0}\Big).
\end{split}
\end{equation} 
Substitute Eq.(\ref{Dsq}) into Eq.(\ref{eq:EOMDirac01}) to obtain 
\begin{equation}\label{eq:EOMConstantFlux}
\begin{split}
0
          &=\varphi''(u)
          +\frac{R'}{R}\varphi'(u)
          +\left[\frac{R''}{2R}
          +\frac{m'\sigma R'-m'^2-\left(\frac{R'}{2}\right)^2}{R^2}+k^2\right] \varphi(u),
\end{split}
\end{equation} 
where $m'=m-\displaystyle{\frac{\phi}{\phi_{0}}}$, and the magnetic flux quantum $\phi_{0}\equiv hc/e$.  We have used the momentum parameter $k^{2} \equiv (E^{2}-M^{2}c^{4})/\hbar^{2}c^{2}$ and $\sigma$ is a spin-state index corresponding to spin up ($\sigma$=+1) or down ($\sigma=-1$).

To be specific, we parametrise the shape of wormhole by setting $R(u)=a\cosh_{q}(u/r)$. They are based on a $q$-deformation of the usual hyperbolic functions defined by~\cite{Arai}
\begin{equation} \label{eq:2}
\cosh_q(x)\equiv\frac{e^{x}+qe^{-x}}{2}, \quad \sinh_q(x)\equiv\frac{e^{x}-qe^{-x}}{2}, \quad \tanh_q(x)=\frac{\sinh_q(x)}{\cosh_q(x)}.
\end{equation}\\
Note the relevant properties
\begin{equation} \label{eq:3}
\cosh_q^2(x)-\sinh^2_q(x)=q, \quad \frac{d}{dx}\sinh_q(x)=\cosh_q(x), \quad \frac{d}{dx}\tanh_q(x)=\frac{q}{\cosh^{2}_q(x)}.
\end{equation}
\\
The deformed functions reduce to hyperbolic functions when $q=1$. For this choice, the wormhole will possess two Hilbert horizons at $u_{p,m}=r\ln \displaystyle{\Big(\sqrt{q+\frac{r^2}{a^2}}\pm \frac{r}{a}\Big)}$ where $R'(u_{p,m})=\pm 1$. Then define a variable $X(u)\equiv rR'(u)/a=\sinh_{q}(u/r)$  to obtain 
\bea\label{eq:EOMX1}
0  &=&\Big(q+X^2\Big)\varphi''(X) +2X\varphi'(X)
         +k^2 r^2\varphi(X)+\Big[\frac{\frac{q}{4}+\frac{r}{a}\sigma m' X-\Big(\frac{r}{a}m'\Big)^2 }{\Big(q+X^2\Big)}
         +\frac{1}{4} \Big]\varphi(X).
\eea   
Define weighting function solution $\varphi(X)=(\sqrt{q}+iX)^\alpha (\sqrt{q}-iX)^\beta\Phi(X)$, the equation of motion can be rewritten as
\begin{equation}\label{eq:EOMX2}
\begin{split}
0
         &=(q+X^2)\Phi''(X) +2\left[(\alpha+\beta+1)X+i(\alpha-\beta)\sqrt{q}\right]\Phi'(X)
        +\left[(\alpha+\beta+\frac{1}{2})^{2}+k^{2}r^{2}\right]\Phi(X),                \\
0
         &=(q+X^2)\Phi''(X) +\Big[\textup{A}+\textup{B}X\Big]\Phi'(X)
         +\Big[\textup{C}+k^2 r^2 \Big]\Phi(X),
\end{split}
\end{equation}
where we assume
\begin{equation}\label{eq:30}
2i\Big(\alpha^2-\beta^2\Big)\sqrt{q}+\frac{r}{a}\sigma m'=0, 
\qquad -2\Big(\alpha^2+\beta^2\Big)q+\frac{q}{4}-\Big(\frac{r}{a}m'\Big)^2=0,  \nonumber
\ee
leading to
\be
\alpha=\kappa_{1}\left(\frac{1}{4}+\frac{i}{\sqrt{q}}\frac{\sigma m'r}{2a}\right), 
\qquad \beta=\kappa_{2}\left(\frac{1}{4}-\frac{i}{\sqrt{q}}\frac{\sigma m'r}{2a}\right), \quad\text{where} \;\;\kappa_{1},\kappa_{2}=\pm 1.
\end{equation}
The coefficients are defined as the following
\begin{equation}\label{eq:31}
\textup{A}
         =2i(\alpha-\beta)\sqrt{q}, \quad\textup{B}=2(\alpha+\beta+1), \quad \textup{C}
         =(\alpha+\beta)(\alpha+\beta+1)+\frac{1}{4}.
\end{equation}
Depending on the sign choices of $\kappa_{1},\kappa_{2}$, the resulting equation of motion and the corresponding energy levels will be dependent or independent of the spin-orbit coupling term $\sim \sigma mr/a\sqrt{q}$.

Define $X=-i\sqrt{q}Y$, the equation then takes the form
\begin{equation}\label{eq:34}
0=(1-Y^2)\Phi''(Y)+2\Big[(\alpha-\beta)-(\alpha+\beta+1)Y\Big]\Phi'(Y)-\Big[(\alpha+\beta)(\alpha+\beta+1)+k^2 r^2+\frac{1}{4}\Big]\Phi(Y)
\end{equation}
Eq.(\ref{eq:34}) is the Jacobi Differential Equation, the energy levels become

\begin{equation}\label{eq:EnergyWormholeConsFlux}
E^{2}_{n,m'}=M^{2}c^{4}+\hbar^{2}c^{2}k^{2}_{n,m'} =M^{2}c^{4}-\frac{\hbar^{2}c^{2}}{r^2}\left(n+\frac{1}{2}+\alpha+\beta \right)^{2}.
\end{equation}

The solutions to Eq.(\ref{eq:34}) are the Jacobi polynomials~\cite{Egrifes}
\begin{equation}\label{eq:37}
\begin{split}
\Phi_{n}(Y)
          &=P^{(\alpha_{0},\beta_{0})}_{n}(Y)=\frac{(-1)^{n}}{2^{n}n!}(1-Y)^{-\alpha_{0}}(1+Y)^{-\beta_{0}}\frac{d^{n}}{dY^{n}}\Big[(1-Y)^{\alpha_{0}}(1+Y)^{\beta_{0}}(1-Y^2)^{n}\Big]
\end{split}
\end{equation}
for integer $n$ and $\alpha_{0}=2\alpha, \beta_{0}=2\beta$. Finally, the solutions of Eq.(\ref{eq:EOMConstantFlux}) is 
\begin{equation}\label{eq:SolutionInsideConFlux01}
\varphi_{n,m',\sigma}(\kappa_{1},\kappa_{2},u)=\left(\sqrt{q}+iX\right)^{\alpha} \left(\sqrt{q}-iX\right)^{\beta} P_{n}^{(2\beta,2\alpha)}\left(iX/\sqrt{q}\right),
\end{equation}
where $X(u)=\sinh_{q}(u/r)$ and $\kappa_{1},\kappa_{2}=\pm 1$.  These are the solutions with QNMs found in Ref.~\cite{Rojjanason:2018icy}. The momentum $k$ for these solutions is generally a complex quantity, so they are not exactly normal modes even when the energy $E$ is real. As will be shown subsequently in \ref{realksect} and \ref{genmSect}, we can analytically continue the solutions to the general $n<0$ cases where the normal modes with real energy and momentum can be found as a special case with $n=-1/2$. Other quasinormal $n<0$ solutions are also relevant in the scattering processes.

\begin{table}[h!]
\centering
\begin{tabular}{ |c|c|c|c|c| } 
 \hline
 $\kappa_{1}$ & $\kappa_{2}$ & $k_{n,m'}r$ & $E_{n,m'}^{2}$ & $\varphi(u)$  \\ 
 
 \hline
 
 &  &  &  &   \\ 

 $+$ & $+$ & $i\left(n+1\right)$ & $M^{2}c^{4}-\frac{\hbar^{2}c^{2}}{r^2}\left(n+1\right)^{2}$ & $\displaystyle{  \left(q+X^{2}\right)^{1/4}\left(\frac{\sqrt{q}+iX}{\sqrt{q}-iX}\right)^{i\frac{\sigma m'r}{2a\sqrt{q}}}P_{n}^{\left(\frac{1}{2}+i\frac{\sigma m'r}{a\sqrt{q}},\frac{1}{2}-i\frac{\sigma m'r}{a\sqrt{q}}\right)}\left(iX/\sqrt{q}\right) }$ \\  
 
  &  &  &  &   \\ 
  
 \hline
 
  &  &  &  &   \\ 
  
 $-$ & $-$ & $in$ & $M^{2}c^{4}-\frac{\hbar^{2}c^{2}}{r^2}n^{2}$  &  $\displaystyle{  \left(q+X^{2}\right)^{-1/4}\left(\frac{\sqrt{q}-iX}{\sqrt{q}+iX}\right)^{i\frac{\sigma m'r}{2a\sqrt{q}}}P_{n}^{\left(-\frac{1}{2}-i\frac{\sigma m'r}{a\sqrt{q}},-\frac{1}{2}+i\frac{\sigma m'r}{a\sqrt{q}}\right)}\left(iX/\sqrt{q}\right) }$ \\ 
 
  &  &  &  &   \\ 
  
 \hline
 
  &  &  &  &   \\ 
  
  $+$ & $-$ &  $i\left(n+\frac{1}{2}\right)-\frac{\sigma m'r}{a\sqrt{q}}$ & $M^{2}c^{4}-\frac{\hbar^{2}c^{2}}{r^2}\left(n+\frac{1}{2}+i\frac{\sigma m'r}{a\sqrt{q}}\right)^{2}$ &  $\displaystyle{  \left(q+X^{2}\right)^{i\frac{\sigma m'r}{2a\sqrt{q}}}\left(\frac{\sqrt{q}+iX}{\sqrt{q}-iX}\right)^{1/4}P_{n}^{\left(\frac{1}{2}+i\frac{\sigma m'r}{a\sqrt{q}},-\frac{1}{2}+i\frac{\sigma m'r}{a\sqrt{q}}\right)}\left(iX/\sqrt{q}\right) }$   \\ 
  
   &  & &  &   \\ 
   
 \hline
 
  &  &  & &   \\ 
  
  $-$ & $+$ &  $i\left(n+\frac{1}{2}\right)+\frac{\sigma m'r}{a\sqrt{q}}$ & $M^{2}c^{4}-\frac{\hbar^{2}c^{2}}{r^2}\left(n+\frac{1}{2}-i\frac{\sigma m'r}{a\sqrt{q}}\right)^{2}$  &  $\displaystyle{  \left(q+X^{2}\right)^{-i\frac{\sigma m'r}{2a\sqrt{q}}}\left(\frac{\sqrt{q}-iX}{\sqrt{q}+iX}\right)^{1/4}P_{n}^{\left(-\frac{1}{2}-i\frac{\sigma m'r}{a\sqrt{q}},\frac{1}{2}-i\frac{\sigma m'r}{a\sqrt{q}}\right)}\left(iX/\sqrt{q}\right) }$   \\ 
  
   &  &  &  &   \\ 
   
 \hline
\end{tabular}
\caption{The wave vector along the $u$ direction $k_{n,m'}$, the energy levels $E_{n,m'}$, and the solutions $\varphi(u)$ of the Dirac equation in two dimensional wormhole with constant magnetic flux through the throat of the wormhole. } \label{Tabel:ConFlux}
\end{table}

\subsection{Solutions in the Upper and Lower plane}

In the flat upper~($R'(u)=1$) and lower~($R'(u)=-1$) plane region outside the wormhole, the equation of motion, Eq.(\ref{eq:EOMConstantFlux}), takes the form
\begin{equation}\label{eq:outsideEq1}
\begin{split}
0=  &\varphi''(u)\pm\frac{1}{R(u)}\varphi'(u)
+\left[k^{2}-\frac{\left(m'\mp\sigma/2\right)^{2}}{(R(u))^{2}}\right]\varphi(u),
\end{split}
\end{equation}
with 
\be \label{Rpeq}
R(u)=\Bigg{\{}\begin{array}{cc}\begin{split}
u-u_{p}+R_{p},\quad\quad  \quad\quad {\rm for}\quad u>u_{p} \\
-(u-u_{m})+R_{m}, \quad\quad {\rm for}\quad u<u_{m},\\
\end{split}\end{array}
\ee
respectively. $R_{p,m}\equiv R(u_{p,m})$ is the corresponding radial distance on the upper and lower plane. Wave solutions of Eq.(\ref{eq:outsideEq1}) are the Hankel function of the first and second kind. Generically the solutions can be expressed as
\begin{equation}\label{eq:outsideSolution1}
\varphi^{(+)}(u)=I_{m,\sigma}H^{(2)}_{m'-\sigma/2}\left(kR(u)\right)+R^{(+)}_{m,\sigma}H^{(1)}_{m'-\sigma/2}\left(kR(u)\right),
\end{equation}
for $u \geq u_{p}$ in the upper plane, and
\begin{equation}\label{eq:outsideSolution2}
\varphi^{(-)}(u)=T_{m,\sigma}H^{(1)}_{m'+\sigma/2}\left(kR(u)\right)+R^{(-)}_{m,\sigma}H^{(2)}_{m'+\sigma/2}\left(kR(u)\right),
\end{equation}
for $u \leq u_{m}$ in the lower plane. The Hankel function of the first~(second) kind corresponds to the waves propagating in $+(-)\hat{u}$ direction respectively.

\section{Matching conditions in the scattering}  \label{secOutWormhole}

The matching conditions of the wave functions are the equality of the complex energy~($E$) and angular momentum~($m'$), and the {\it smooth} continuity~(i.e., $C^{1}$ continuity) of the wave functions between the inner and outer region of the wormhole. The momentum $k$ will also be equal due to the relation (\ref{eq:EnergyWormholeConsFlux}).

Consider the incoming waves propagating in the upper plane region scatter with the wormhole.  At the upper Hilbert horizon $u_{p}$, the waves will be partially reflected back and partially transmitted into the inner region of the wormhole. The transmitted waves will be again partially reflected and partially transmitted into the lower plane region at the lower Hilbert horizon $u_{m}$.  In this scenario, there is no incoming waves at the lower Hilbert horizon, i.e., $R^{(-)}_{m,\sigma}=0$. We can normalize $I_{m,\sigma} \equiv 1$ and $R^{(+)}_{m,\sigma}\equiv R_{m,\sigma}$ where the Hankel function of the first and second kind correspond to outgoing and incoming waves, respectively. In the wormhole region $u_{m}<u<u_{p}$, the general solution from Eq.(\ref{eq:SolutionInsideConFlux01}) can be expressed as

\begin{equation}\label{eq:SolutionInside02}
\begin{split}
\varphi^{(in)}(u)=
&\;
    \varphi_{m,n,\sigma}(+,-,u)+\varphi_{m,n',\sigma'}(-,+,u)\\=
&
     A_{m,n,\sigma}\left(\sqrt{q}+iX\right)^{\alpha_{0}} \left(\sqrt{q}-iX\right)^{-\alpha_{0}^{*}} P_{n}^{(-2\alpha_{0}^{*},2\alpha_{0})}\left(iX/\sqrt{q}\right) \\
&
    + B_{m,n',\sigma'}\left(\sqrt{q}+iX\right)^{-\alpha_{0}'} \left(\sqrt{q}-iX\right)^{\alpha_{0}'^{*}} P_{n'}^{(2\alpha_{0}'^{*},-2\alpha_{0}')}\left(iX/\sqrt{q}\right),
\end{split}\end{equation}
where $X=\sinh_{q}(u/r),\;\; \alpha=\kappa_{1}\alpha_{0},\;\;\beta=\kappa_{2}\alpha_{0}^{*}$, and $ \alpha_{0}\equiv\frac{1}{4}+\frac{i}{\sqrt{q}}\frac{\sigma m'r}{2a}$ and the condition $E_{n,\sigma}=E_{n',\sigma'}$ is required. Note that the parameter $\alpha_{0}$ also depends on $m,\sigma$.

\begin{figure}[H]
  \centering
  \includegraphics[width=0.6\textwidth]{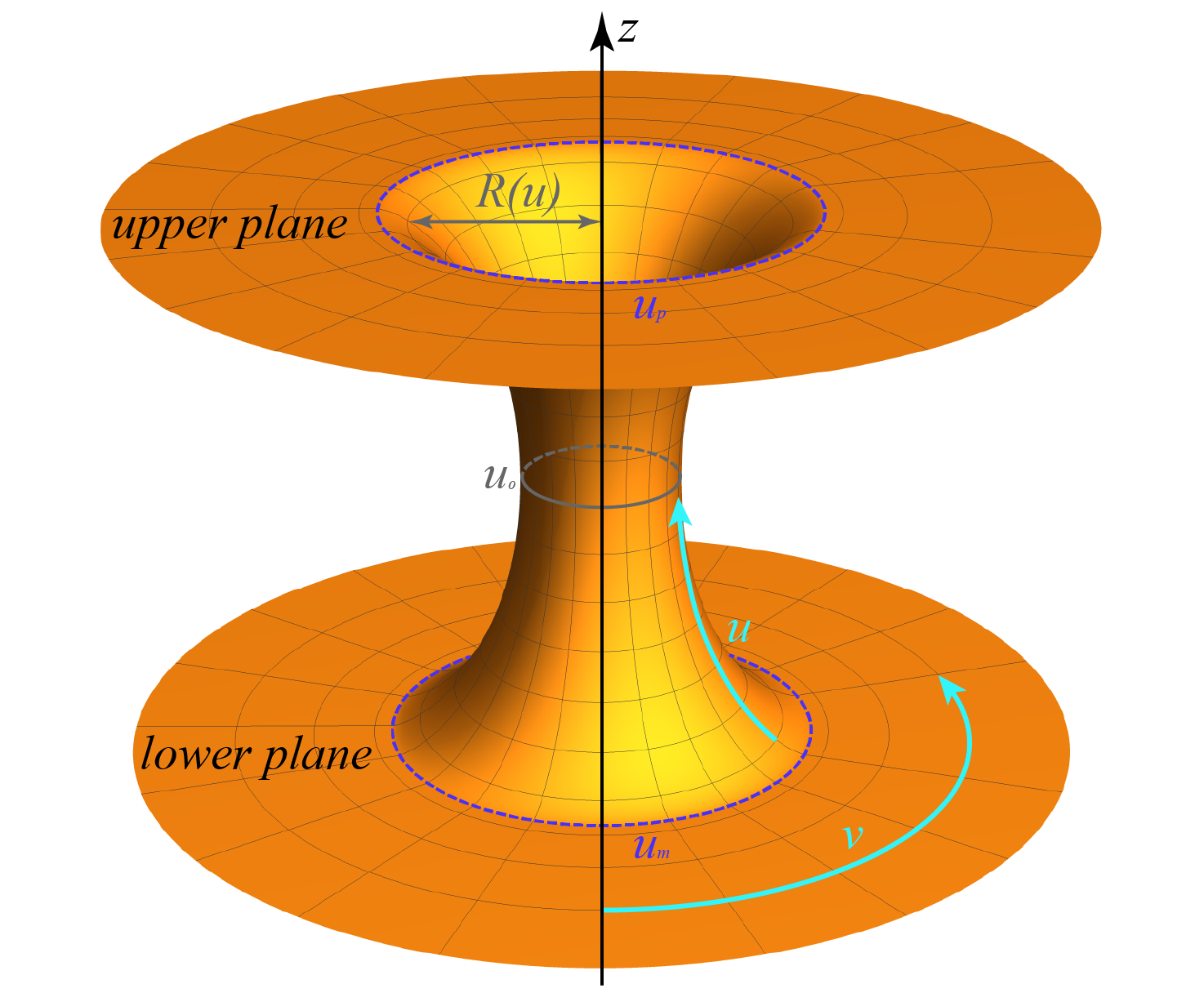}
  \caption{A wormhole connected smoothly to two flat planes at Hilbert horizons $u_{p}$ and $u_{m}$, the midpoint of wormhole is at $u_{0}=\frac{1}{2}\ln q$ where $R(u)$ is minimum. The wormhole is symmetric with respect to $u_{0}$.}
 \label{fig2}
\end{figure}

\subsubsection{ The matching condition : At upper surface\label{secUpperBC}}

The boundary between the inner and outer region of the wormhole is at $R(u_{p})\equiv R_{p}$, see Fig.~\ref{fig2}. The wave function and the first derivative of the wave function must be continuous at the boundary. From conservation law of energy-momentum, $E^{(in)}=E^{(out)}$ and $ k^{(in)}= k^{(out)}$. The first boundary condition at the upper plane is
 \begin{equation}\label{eq:BC01}
 \begin{split}
\varphi^{(+)}(u_{p})
&
    =\varphi^{(in)}(u_{p})    \\
H^{(2)}_{m'-\sigma/2}\left(kR_{p}\right)+R_{m,\sigma}H^{(1)}_{m'-\sigma/2}\left(kR_{p}\right)
&
    =A_{m,n,\sigma}\left(\sqrt{q}+iX\right)^{\alpha_{0}} \left(\sqrt{q}-iX\right)^{-\alpha_{0}^{*}} P_{n}^{(-2\alpha_{0}^{*},2\alpha_{0})}\left(iX/\sqrt{q}\right) \\
&\quad
    + B_{m,n,\sigma}\left(\sqrt{q}+iX\right)^{-\alpha_{0}} \left(\sqrt{q}-iX\right)^{\alpha_{0}^{*}} P_{n}^{(2\alpha_{0}^{*},-2\alpha_{0})}\left(iX/\sqrt{q}\right),
\end{split}
\end{equation}
where $\sinh_{q}\left(u_{p}/r\right)=r/a$, and $R_{p}=\sqrt{qa^{2}+r^{2}}$. The derivative of the wave function at the boundary $u=u_{p}$
\begin{equation}\label{eq:BC02}
 \begin{split}
\frac{d\varphi^{(+)}(u)}{du}|_{u_{p}}&=k\left(H^{(2)}_{m'-\sigma/2}{}'\left(kR_{p}\right)+R_{m,n}H^{(1)}_{m'-\sigma/2}{}'\left(kR_{p}\right)\right).
\end{split}
\end{equation}

\subsubsection{ The matching condition : At lower surface\label{secLowerBC}}

Even for general $q\geq 0$, the wormhole is symmetric with respect to the midpoint $u_{0}=\ln q/2$. The Hilbert horizon at the lower surface is at $u_{m}$ where $R(u_{m})\equiv R_{m}=R_{p}$. The second boundary condition takes the form
 \begin{equation}\label{eq:BC03}
 \begin{split}
\varphi^{(-)}(u_{m})
&
    =\varphi^{(in)}(u_{m})    \\
T_{m,\sigma}H^{(1)}_{m'+\sigma/2}\left(kR_{m}\right)
&
    =A_{m,n,\sigma}\left(\sqrt{q}-iX\right)^{\alpha_{0}} \left(\sqrt{q}+iX\right)^{-\alpha_{0}^{*}} P_{n}^{(-2\alpha_{0}^{*},2\alpha_{0})}\left(-iX/\sqrt{q}\right) \\
&\quad
    + B_{m,n,\sigma}\left(\sqrt{q}-iX\right)^{-\alpha_{0}} \left(\sqrt{q}+iX\right)^{\alpha_{0}^{*}} P_{n}^{(2\alpha_{0}^{*},-2\alpha_{0})}\left(-iX/\sqrt{q}\right).
\end{split}
\end{equation}
The derivative of the wave function at the second boundary $u=u_{m}, R=R_{m}$ is
\begin{equation}\label{eq:BC04}
 \begin{split}
\frac{d\varphi^{(-)}(u)}{du}|_{u_{m}}  &=-kT_{m,\sigma}H^{(1)}_{m'+\sigma/2}{}'\left(kR_{m}\right).
\end{split}
\end{equation}
From the matching conditions (\ref{eq:BC01})-(\ref{eq:BC04}), the scattering coefficients $A_{m,n,\sigma},B_{m,n,\sigma},T_{m,\sigma},R_{m,n}$ can be solved for which here and henceforth the subscripts will be suppressed.

\section{Analytic relation between Reflection and Transmission coefficients}  \label{secgenrel}

In this section a very general analytic relation between the reflection and transmission coefficients are derived from the equation of motions. The relation can be expressed in terms of Wronskian of the solutions in each region.  The resulting relation is then verified numerically for each scattering scenario. 

The generic form of the equation of motion can be expressed as
\bea
0&=&\varphi''(u)+f(u)\varphi'(u)+g(u)\varphi(u), \label{geom}
\eea
where 
\be
f(u)=\frac{R'}{R}=\Bigg{\{} 
\begin{array}{cc}\begin{split}
\mp\frac{1}{R(u)},\quad\quad  \quad\quad {\rm for}\quad u<u_{m}, u>u_{p} \\
\frac{\tanh_{q}\left( \frac{u}{r}\right)}{r}, \quad\quad {\rm for}\quad u_{m}<u<u_{p},\\
\end{split}\end{array}
\ee
and
\be
g(u)=\left[\frac{R''}{2R}
          +\frac{m'\sigma R'-m'^2-\left(\frac{R'}{2}\right)^2}{R^2}+k^2\right]=\Bigg{\{} \begin{array}{cc}\begin{split}
\left[\frac{\mp m'\sigma -m'^2-\frac{1}{4}}{R(u)^2}+k^2\right],\quad\quad \quad\quad\quad {\rm for}\quad u<u_{m}, u>u_{p} \quad\quad\\
\left[\frac{1}{2r^{2}}
          -\frac{\left(m'-\frac{a\sigma}{2r}\sinh_{q}(\frac{u}{r})\right)^2}{a^{2}\cosh_{q}^{2}(\frac{u}{r})}+k^2\right], \quad\quad {\rm for}\quad u_{m}<u<u_{p},\quad\quad\\
\end{split}\end{array}
\ee
respectively. Note that $R(u)$ in the planes is given by (\ref{Rpeq}). For any two solutions $\varphi_{1},\varphi_{2}$ satisfying (\ref{geom}), by multiplying the equation of motion of one solution with the other solution then subtracting the two equations we obtain
\bea
\varphi_{1}\varphi''_{2}-\varphi_{2}\varphi''_{1}+f(u)(\varphi_{1}\varphi'_{2}-\varphi_{2}\varphi'_{1})&=&0,
\eea
or in terms of the Wronskian $W_{12}\equiv \varphi_{1}\varphi'_{2}-\varphi_{2}\varphi'_{1}$,
\bea
\frac{W_{12}'}{W_{12}}+f(u)&=&0.
\eea
By integrating this equation in the region covering the wormhole, 
\be
\ln\left(\frac{W(u_{p})}{W(u_{m})}\right)= -\int_{u_{m}}^{u_{p}}f(u)~du,  \label{grel}
\ee
is the resulting general relation and the subscript of the Wronskian has been omitted since the relation is valid for any pairs of solutions.  The RHS of (\ref{grel}) can be calculated explicitly,
\bea
-\int_{u_{m}}^{u_{p}}f(u)~du&=&\ln\left(\frac{\cosh_{q}\left(\frac{u_{m}}{r}\right)}{\cosh_{q}\left(\frac{u_{p}}{r}\right)}\right).  \label{anrel}
\eea
Since the wormhole is symmetric with $R(u_{m})=R(u_{p})=\sqrt{r^{2}+qa^{2}}$, the Wronskian at the two Hilbert horizons are always equal, $W(u_{m})=W(u_{p})$.  

The general relation (\ref{grel}) can be applied to the two solutions $\varphi$ and $\varphi^{*}$, specifically at $u_{m},u_{p}$ where the wave functions and their first derivatives in the two connecting regions are equal.  In terms of the solutions in the outer regions, we have the relation~(with the subscripts suppressed)
\bea
-|T|^{2}\left(H^{(2)}(kR_{m})H^{(2)*'}(kR_{m})-{\rm cc.}\right)&=&\left(H^{(2)}(kR_{p})H^{(2)*'}(kR_{p})-{\rm cc.}\right) + |R|^{2}\left(H^{(1)}(kR_{p})H^{(1)*'}(kR_{p})-{\rm cc.}\right) \notag \\
&&+R\left(H^{(1)}(kR_{p})H^{(2)*'}(kR_{p})-H^{(1)'}(kR_{p})H^{(2)*}(kR_{p})\right)-{\rm cc.}\label{TRrel}
\eea

\section{Scattering for normal and quasi-normal modes}  \label{secScat}

In this section we consider scattering of fermion in 2-dimensional planar surface with $(1+2)$-dimensional wormhole. Even this ``wormhole'' is not the actual spacetime wormhole as in GR and other gravity theories where time dilatation exists, its scattering still reveals a number of interesting properties. As discussed in details below, scattering of physical fermion with real momentum and energy has quantized behaviour that relates momentum and angular momentum of the scattered fermion.  For scattering involving QNMs of the fermion, enhancement~(and attenuation) of scattered fermion is a possibility. The wormhole and spin parameters are set to $r=a=q=1=\sigma$ for all of the numerical results in this section.

\subsection{scattering for real momentum $k$}  \label{realksect}

Naturally, incoming waves from outside region are generated with real momentum and energy.  When scattered with the wormhole, only real momentum and energy states~(i.e., normal modes) will be allowed to propagate through the hole and transmitted to the other outside region.  The scattering waves will be partially reflected back and partially transmitted through the hole to the other side. From Eqn.~(\ref{eq:EnergyWormholeConsFlux}), the only possibility for real positive momentum $k$~(in units of $\hbar c$) is when $\alpha=-\beta^{*}=\pm \displaystyle{\left(\frac{1}{4}+\frac{i\sigma m' r}{2a\sqrt{q}}\right)}$ and $n=-1/2$. This is the analytic continuation of the equation of motion (\ref{eq:34})~(with (\ref{eq:EnergyWormholeConsFlux}) substituted) to $n<0$ cases. The momentum then takes the value 
\be
k = \frac{m'}{a\sqrt{q}}\equiv k_{m'},  \label{kmrel}
\ee
notably a {\it momentum-angular momentum relation}. Only the waves with quantized momentum $k_{m'}$ and energy given by $E=\sqrt{M^{2}c^{4}+\hbar^{2}c^{2}k_{m'}^{2}}$ are allowed to pass through the wormhole to the other side. For real $k$ since $H_{\nu}^{(2)*}(kR)=H_{\nu}^{(1)}(kR)$, and using the identity $H_{\nu}^{(1)'}\!(z)H_{\nu}^{(2)}\!(z)-H_{\nu}^{(2)'}\!(z)H_{\nu}^{(1)}\!(z)=4i/\pi z$, the relation between transmission and reflection coefficients from Eqn.~(\ref{TRrel}) is simplified to
\be
|T|^{2}+|R|^{2}=1,  \label{urel}
\ee
the unitarity condition.  Unitarity is the consequence of reality of momentum in the scattering. Table~\ref{Realk} shows scattering coefficients for $m'=1,2,3,4$ cases, all of which the unitarity relation (\ref{urel}) is numerically verified. Wave function profile of the matching is shown in Fig.~\ref{realkfig}.

\begin{table}[h!]

\centering

\begin{tabular}{ |c|c|c|c|c| }

\hline

$m'$ & $A$ & $B$ & $T$ & $R$ \\ \hline

1 & $ 0.486599 + 0.867254 i$   & $ 0.0131312 + 0.0626096 i$  & $ -0.292892 + 0.699874 i$     & $ 0.600951 + 0.251493 i$ \\ \hline

2 & $0.434897 + 0.900917 i$     & $-0.0245842 + 0.0137852 i$ & $-0.408684 + 0.576578 i$       & $ 0.577195 + 0.409121 i$ \\ \hline
  
3 & $0.334517 + 0.942154 i$ & $ -0.0147694 - 0.00833657 i$ & $ -0.516515 + 0.462697 i$ & $ 0.480745 + 0.536663 i$ \\ \hline

4 & $0.219563 + 0.975395 i$ & $ -0.00110507 - 0.0115601 i $ & $ -0.612787 + 0.328918 i$ & $0.339825 + 0.633107 i$ \\ \hline

\end{tabular}

\caption{Scattering coefficients for real momentum scenario.} \label{Realk}

\end{table}

The $m'$-dependence of the transmission coefficient $T$ and $R$ are shown in Fig.~\ref{TRmfig}. Interestingly, $T(m')$ and $R(m')$ converge to oscillating functions for large $m'$. The unitarity condition (\ref{urel}) is always obeyed.

\begin{figure}[ht]
  \centering
   \includegraphics[width=0.45\textwidth]{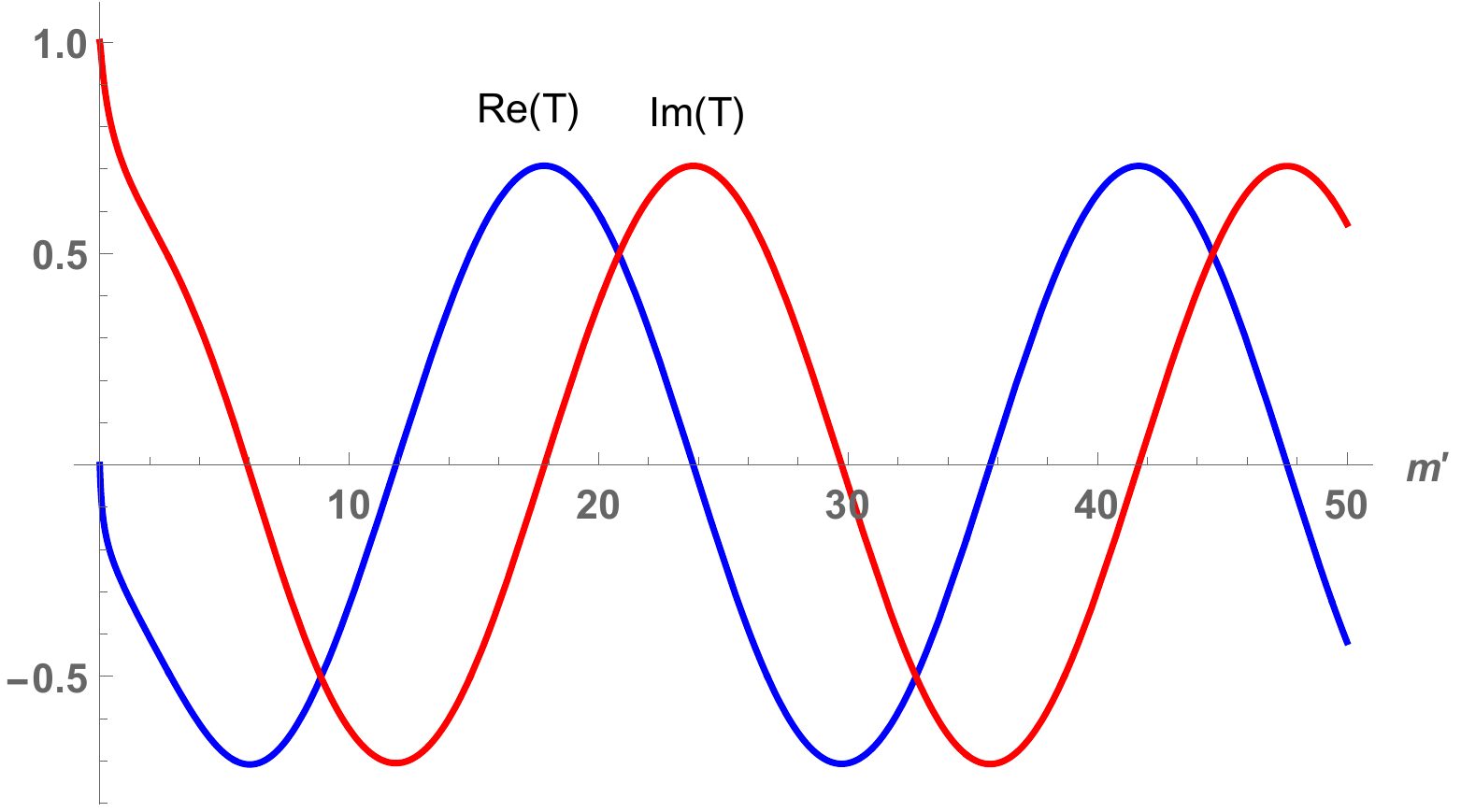}
   \includegraphics[width=0.45\textwidth]{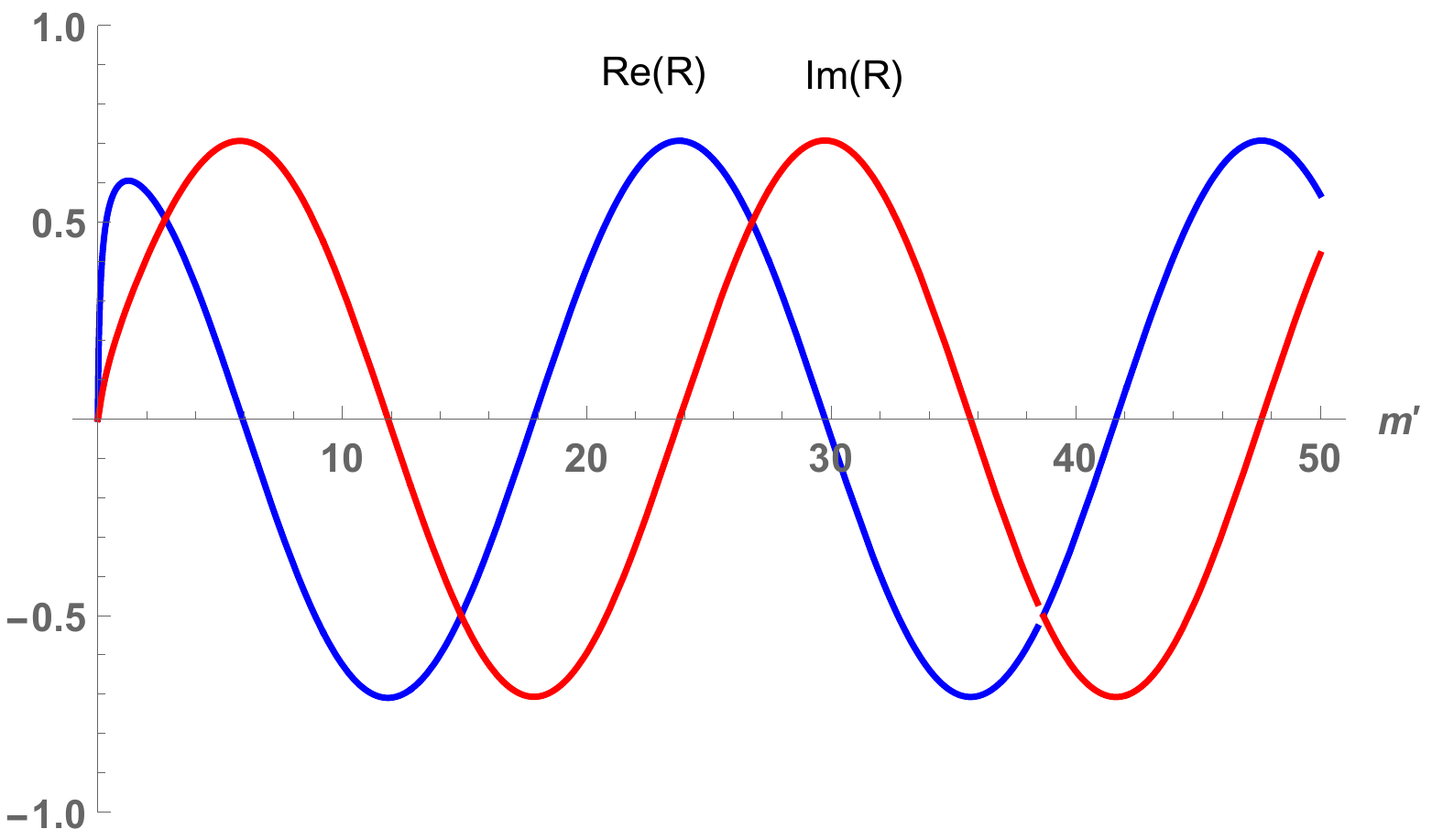}
  \caption{The $m'$-dependence of the transmission coefficient $T$ and reflection coefficient $R$ in the scattering with real momentum scenario.}
\label{TRmfig}
\end{figure}

\subsection{Scattering involving QNMs}

Scattering involving QNMs usually violates unitarity since the waves are decaying in time. It will satisfy a more general relation given by (\ref{TRrel}). There are three possible scenarios for such scattering.
\begin{enumerate}
\item
scattering with $m'=0$
\item
scattering with nonzero $m'$
\item
scattering with spin-flip $\sigma \leftrightarrow -\sigma$ in the inner region
\end{enumerate}

There are three scenarios for scattering involving QNMs of the wormhole. The matching condition requires the energy and inevitably the momentum $k$ to be the same value for inner and outer regions of the wormhole. Generically since the momentum given by (\ref{eq:EnergyWormholeConsFlux}) is always a complex quantity with an exception of $k_{m'}$ discussed in Sect.~\ref{realksect}, the momentum of wave functions in the outer regions will also need to be complex as well.  Consequently, there are {\it spatial} attenuation and enhancement of waves in the outer regions as the waves are concurrently decaying in time. The {\it physical} energy to be measured in experiments is the real part of the (complex) energy used in the matching conditions.

\subsubsection{scattering with $m'=0$}

The scattering in this scenario corresponds to the states in the wormhole with quantized flux $\phi = m \phi_{0}, m=1,2,3,...$ and the orbital angular momentum of the states is $m\hbar$. The quantized magnetic flux is required specifically for the scattering to occur.  In this case the energy could be real for massive fermion and small $n$. On the other hand, the momentum becomes $k=i\displaystyle{\left(n+\frac{1}{2}\right)/r}$, purely imaginary. The waves in the outer regions to be matched with the ones inside the wormhole thus need to be attenuating or enhancing along the $u$ direction. Since the source of the fermion can locate at finite distance from the hole in experimental situation, this scattering scenario is still physically relavant. 

\begin{table}[h!]

\centering

\begin{tabular}{ |c|c|c|c|c| }

\hline

$n$ & $A$ & $B$ & $T$ & $R$ \\ \hline

0 & $ -1.04147 i$ & $ 1.04147 $ & $ 1.70376 i $ & $ 0 $ \\ \hline

1 & $-2.04891$ & $ -2.04891 i $ & $ 4.94571 i$ & $ 0$ \\ \hline

2 & $3.60535 i$ & $-3.60535 $ & $14.3565 i$ & $ 0$ \\ \hline

3 & $ 6.2298 $ & $6.2298 i$ & $ 41.6741 i$ & $ 0$ \\ \hline

\end{tabular}

\caption{Scattering coefficients for $m'=0$ scenario with $r=a=q=1=\sigma$.} \label{m0tab}
\end{table}

The fermionic states inside the wormhole consist of two waves with $n=n'$, $\varphi(+,-,n)$ and $\varphi(-,+,n)$ where $(\kappa_{1},\kappa_{2})=(+,-),(-,+)$ represents $+\hat{u},-\hat{u}$ going modes respectively. The scattering coefficients for $n=0,1,2,3$ are presented in Table~\ref{m0tab}. The transmission and reflection coefficients obey general relation (\ref{TRrel}) but not the unitarity condition (\ref{urel}) due to the complexity of momentum $k$. Remarkably, the waves simply tunnel through the wormhole with no reflection, i.e., $R=0$. The scattering coefficients for $n=0,1,2,3$ are given in Table~\ref{m0tab}. Wave function profile of the matching is shown in Fig.~\ref{m0fig}.

\subsubsection{scattering with nonzero $m'$}   \label{genmSect}

For $m'\neq 0$ the waves inside the wormhole consist of $\varphi(+,-,n)$ and $\varphi(-,+,n')$. According to the energy formula (\ref{eq:EnergyWormholeConsFlux}), the $-\hat{u}$-travelling wave $\varphi(-,+,n')$ must have the wave function associated with $n'=-n-1$ in order to have the same energy, $E_{n}=E_{n'}$.  Since the Jacobi polynomials $P_{n}$ are zero for non-positive integer $n$, we need a more general form of the solutions. From the equation of motion (\ref{eq:34}) with (\ref{eq:EnergyWormholeConsFlux}) substituted,
\be
0=\left(1-Y^2\right) \Phi ''(Y)+(-2\alpha +2\beta +Y (-(2\alpha +2\beta +2))) \Phi '(Y)-n (2\alpha +2\beta +n+1)\Phi (Y) ,
\ee
we can analytically continue the solution to the case where $n<0$ by solving this equation as the hypergeometric equation. There are two solutions,
\be
2^{2\alpha }(Y-1)^{-2\alpha } \, _2F_1\left(\frac{1}{2}-\alpha+\beta-\frac{1}{2} \gamma,\frac{1}{2}-\alpha +\beta+\frac{1}{2} \gamma;1-2\alpha ;\frac{1-Y}{2}\right),
\ee
and
\be
\, _2F_1\left(\frac{1}{2}+\alpha + \beta -\frac{1}{2} \gamma,\frac{1}{2}+\alpha+\beta+\frac{1}{2}\gamma;1+2\alpha;\frac{1-Y}{2}\right),  \label{2sol}
\ee
where
\be
\gamma = \sqrt{-4 n(n+ 2\alpha + 2\beta +1)+(2\alpha+2\beta +1)^{2}}.
\ee
The first solution has kinks or sharp turns in the inner region of wormhole while the second solution is smooth throughout. Therefore we use only the second solution (\ref{2sol}) in the scattering analysis. 

The incoming waves are scattered by the wormhole, thus excite the fermionic modes in the process.  The reflected outcomes occurred at both Hilbert horizons where $u_{p}$ and $u_{m}$. At the same time, some of the waves in the inner region leaks out through both horizons into the outer planes. The leaking waves are attenuated or enhanced depending on the sign of ${\rm Im}(k)$.  For the $-\hat{u}~(+\hat{u})$ moving $H^{(2)}_{\nu}~(H^{(1)}_{\nu})$, the wave function is spatially enhanced~(attenuated) with respect to the increase of $u$ respectively. Wave function profile of the matching is shown in Fig.~\ref{genmfig}. The scattering coefficients for certain parameters are given in Table~\ref{genmtab}. The $m'$-dependence of the transmission coefficient $T$ and $R$ are shown in Fig.~\ref{TRmgenfig}. Remarkably, both $T$ and $R$ show a peak at $m'=\pm 6$. 

\begin{table}[h!]

\centering

\begin{tabular}{ |c|c|c|c|c| }

\hline

$n$ & $A$ & $B$ & $T$ & $R$ \\ \hline
 
0  &  $ -0.955236 + 0.304246 i $  &  $ 0.357174 + 0.524487 i $     & $ -0.946488 + 0.27418 i $    &  $ 2.178 + 4.25366 i $ \\ \hline

1  &  $ -0.517527-0.00380051 i $ &  $ -0.0447471 + 0.569521 i $   & $ -0.695048 - 4.20998 i $    &  $ 20.8443 + 62.0839 i $ \\ \hline
   
2  &  $ -0.436724 - 0.524891 i $  &  $ -0.304297 + 0.350316 i $     & $ 67.6094 - 33.8419 i $       &  $ 408.353 + 1083.51 i $ \\ \hline
 
3  &  $ 0.124303 - 0.795473 i $   &  $ -0.382871 + 0.0366021 i $    & $ 409.066 + 930.761 i $      &  $ 7201.99 + 18766.6 i $ \\ \hline

\end{tabular}

\caption{Scattering coefficients for the wormhole states with $E_{n}=E_{n'}$ where $n'= -n-1, m'=1$.} \label{genmtab}

\end{table}

\begin{figure}[ht]
  \centering
\subfigure[]{%
\includegraphics[width=0.45\textwidth]{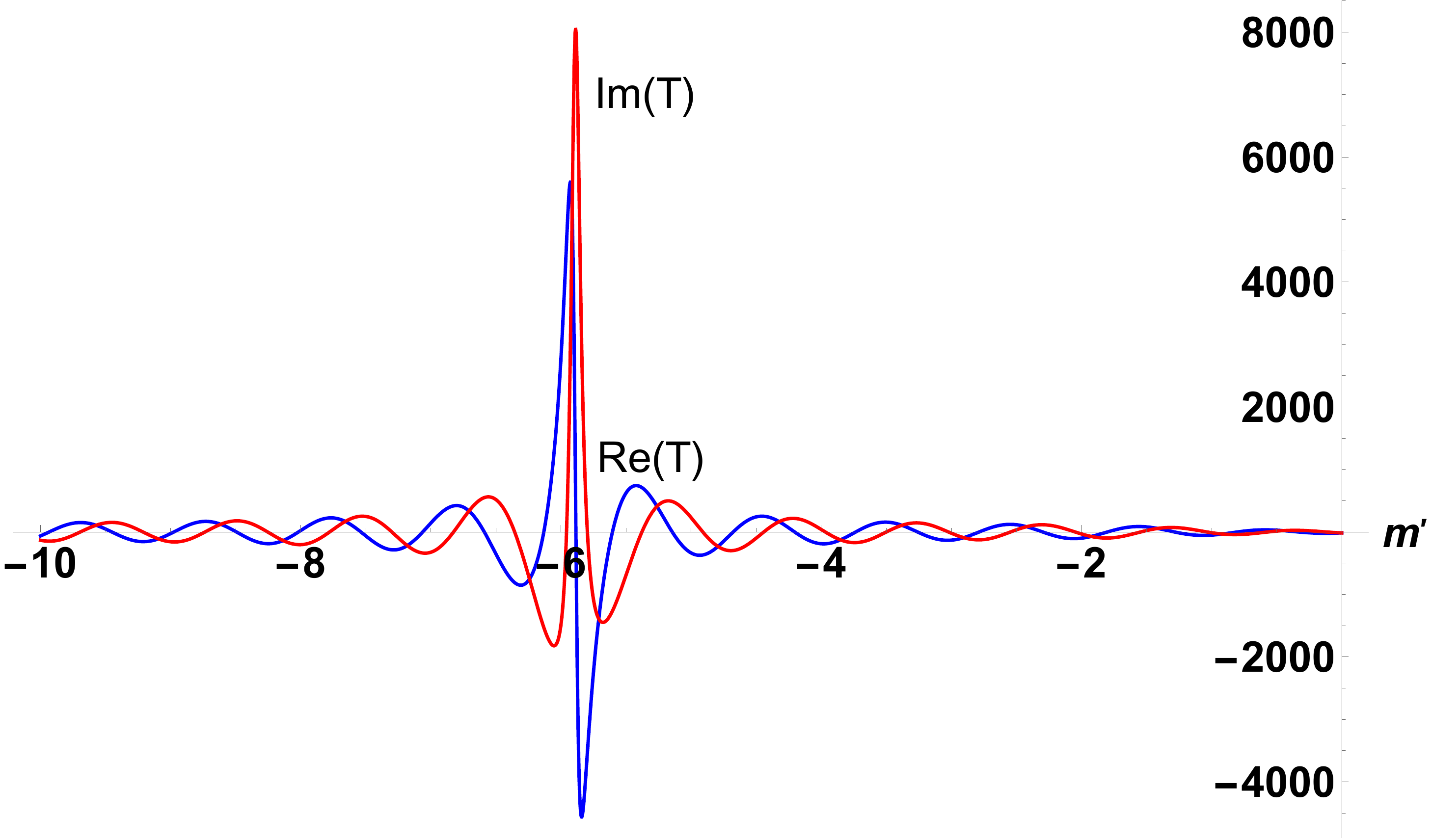}}
\quad
\subfigure[]{%
\includegraphics[width=0.45\textwidth]{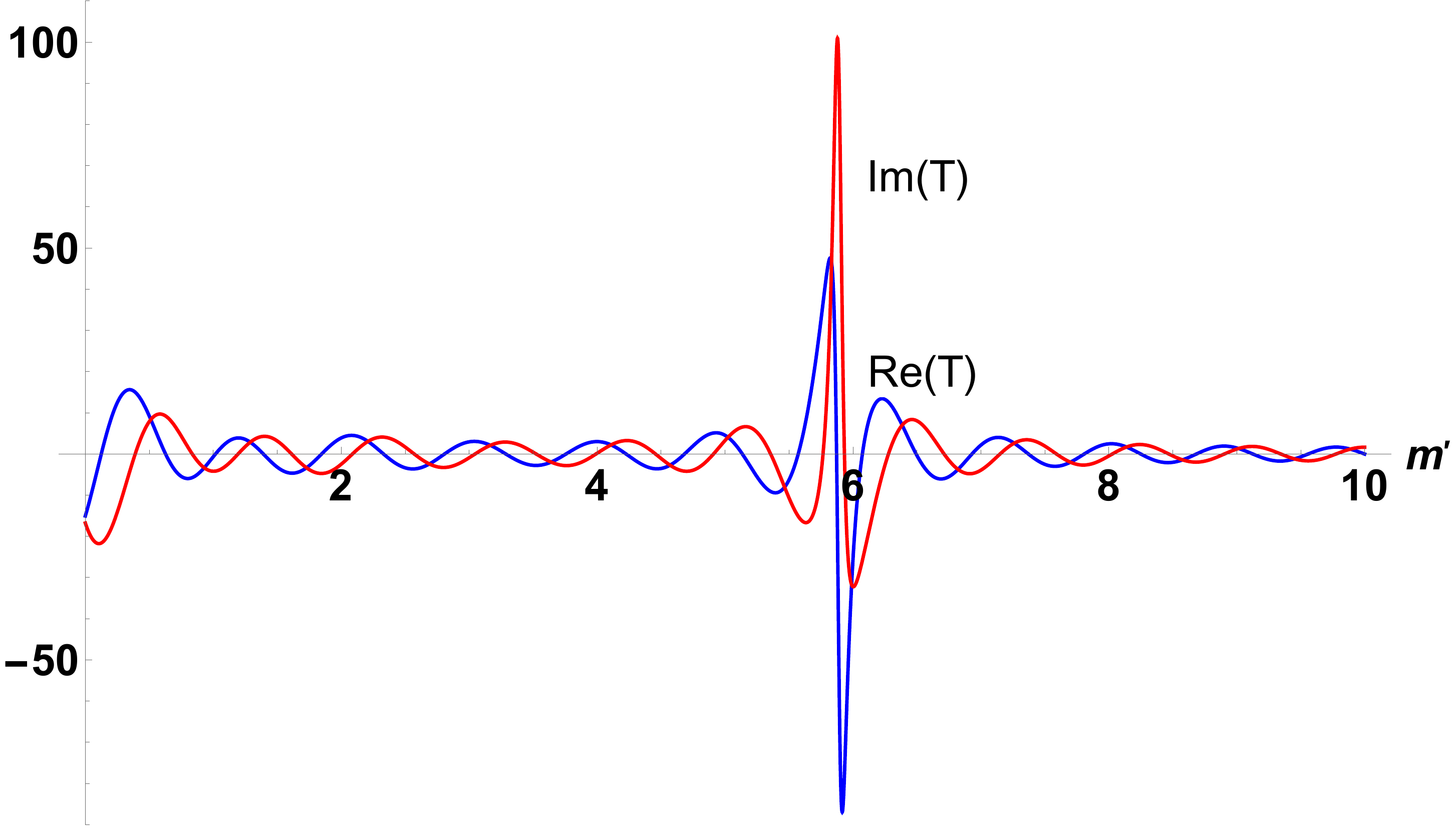}}
\subfigure[]{%
\includegraphics[width=0.45\textwidth]{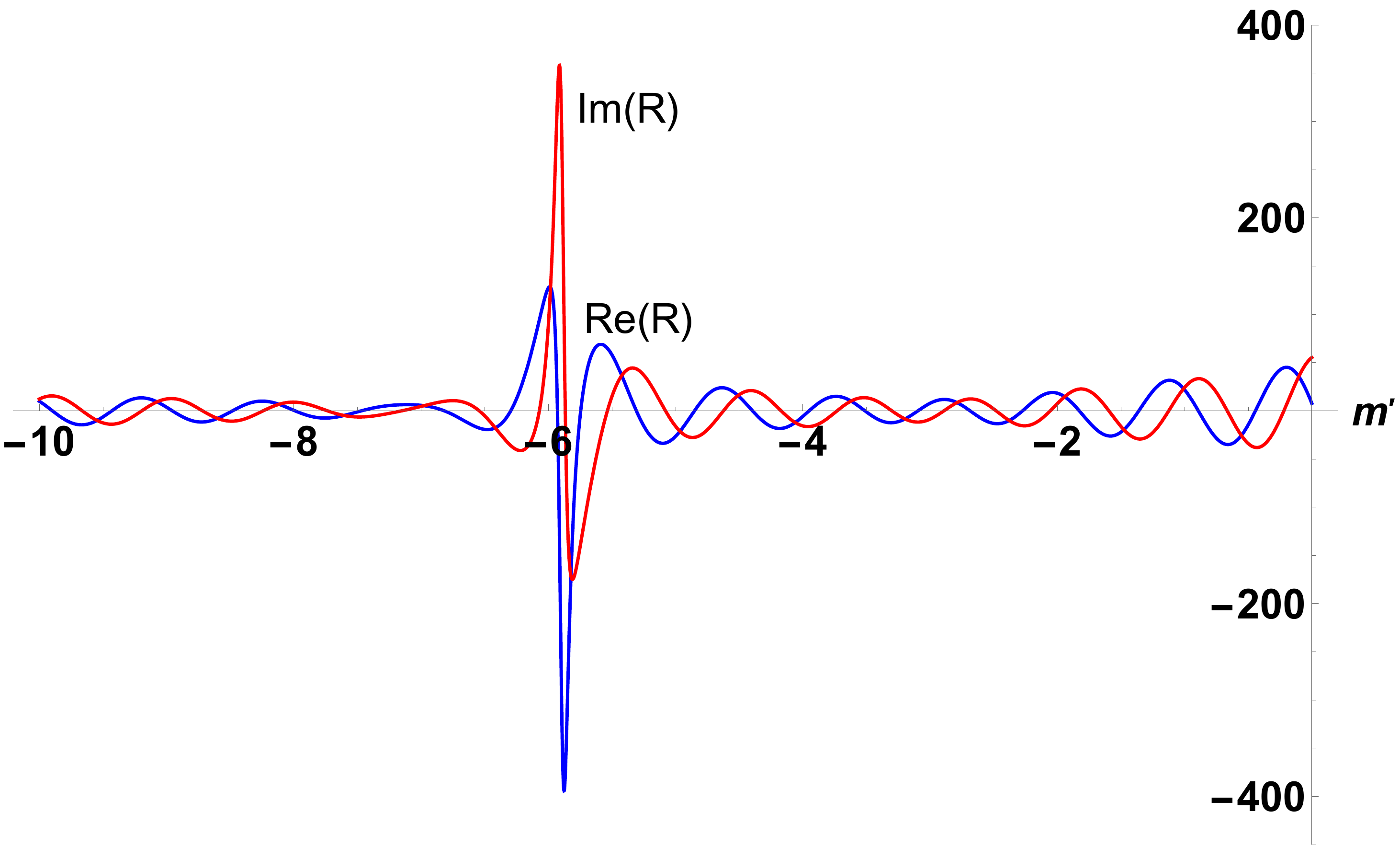}}
\quad
\subfigure[]{%
\includegraphics[width=0.45\textwidth]{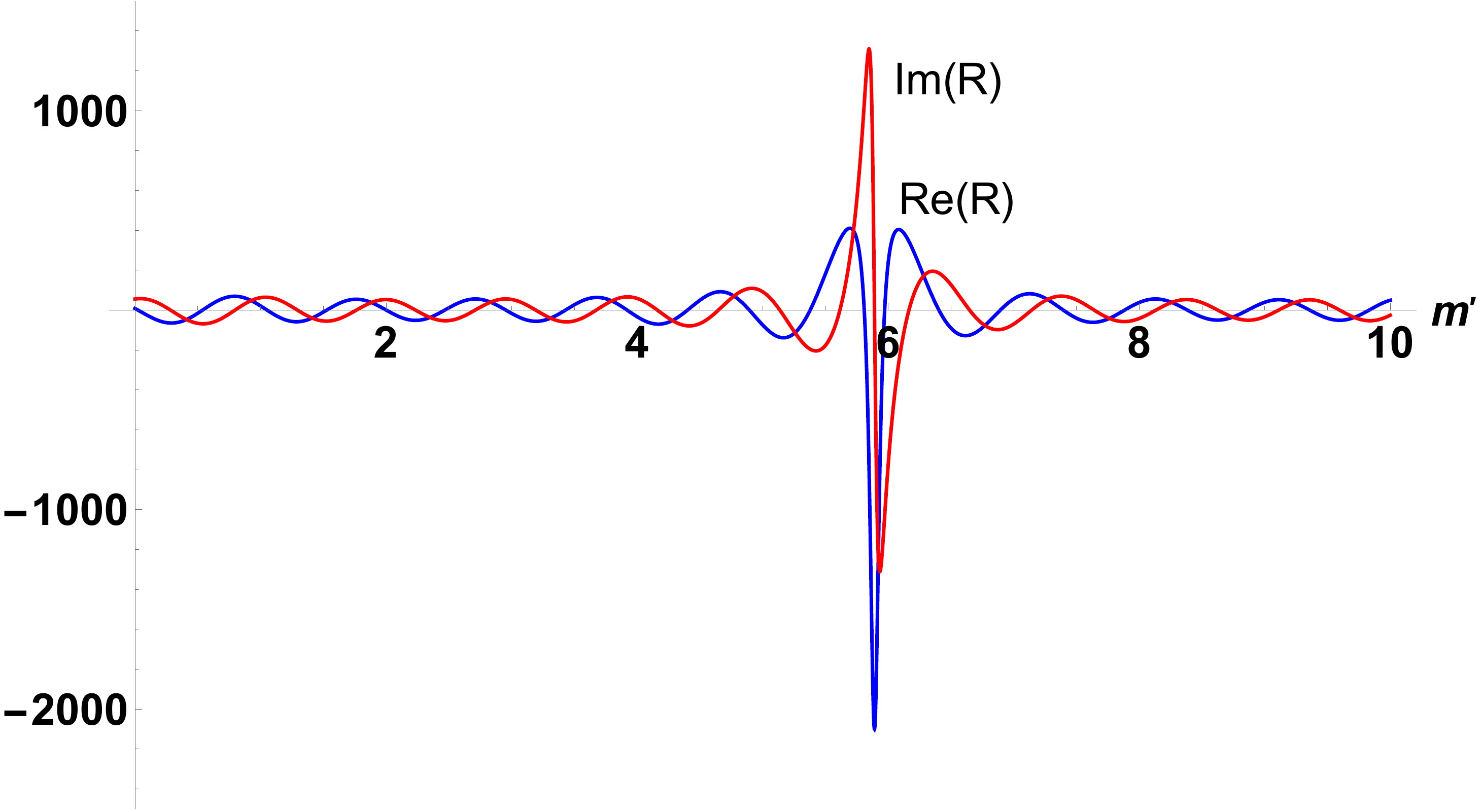}}

\caption{The $m'$-dependence of the transmission coefficient $T$ and reflection coefficient $R$ in the scattering with nonzero $m'$ scenario.}
\label{TRmgenfig}
\end{figure}

Scattering of waves to the wormhole with QNMs excitation is analogous to interaction of oscillating free springs with damped spring.  The springs in all regions will eventually decay to zero due to the imaginary part of QNMs in the factor $e^{-iEt}$ for ${\rm Im}(E)<0$. For QNMs with ${\rm Im}(E)>0$, the system will be unstable, the backreaction to the curve surface from the fermion will be large.

\subsubsection{scattering with spin flip $\sigma \leftrightarrow -\sigma$ in the inner region}

An interesting scenario is the spin-flip scattering while maintaining the same energy. Instead of the reflected waves with different $n$'s as in Section~\ref{genmSect}, the reflected waves in the inner region of the wormhole could get spin-flipped $\sigma \to -\sigma$ with $(\alpha,\beta)\to -(\alpha,\beta)$ and still have the same energy and momentum. From Eqn.~(\ref{eq:SolutionInside02}), this corresponds to $n'=n, \sigma'= -\sigma$. The scattering coefficients for $m'=0,1$ are given in Table~\ref{sftab1},\ref{sftab2} respectively. Note the exact values of the scattering coefficients in $m'=0$ case to the values in Table~\ref{m0tab}. Wave function profile of the matching is shown in Fig.~\ref{Spinflipfig}. The $m'$-dependence of the transmission and reflection coefficients are shown in Fig.~\ref{TRmSFfig}.

\begin{table}[h!]

\centering

\begin{tabular}{ |c|c|c|c|c| }

\hline

$n$ & $A$ & $B$ & $T$ & $R$ \\ \hline

0 & $ -1.04147 i$ & $ 1.04147 $ & $ 1.70376 i $ & $ 0 $ \\ \hline

1 & $-2.04891$ & $ -2.04891 i $ & $ 4.94571 i$ & $ 0$ \\ \hline

2 & $3.60535 i$ & $-3.60535 $ & $14.3565 i$ & $ 0$ \\ \hline

3 & $ 6.2298 $ & $6.2298 i$ & $ 41.6741 i$ & $ 0$ \\ \hline

\end{tabular}

\caption{Scattering coefficients for the spin-flip $\sigma \leftrightarrow -\sigma $ scenario where $m'=0$.} \label{sftab1}

\end{table}

\begin{table}[h!]

\centering

\begin{tabular}{ |c|c|c|c|c| }

\hline

$n$ & $A$ & $B$ & $T$ & $R$ \\ \hline

0 & $0.893023 + 0.432957 i$ & $ -0.618941 + 0.390062 i $ & $ 0.421126 + 0.610085 i $ & $ 0.331071 + 0.331983 i $ \\ \hline

1 & $0.115499 - 1.60794 i$ & $1.11401 + 0.323024 i $ & $ 3.55076 - 1.43318 i$ & $ -0.385585 + 2.55466 i$ \\ \hline

2 & $-2.98693 + 0.177235 i$ & $0.0839698 - 2.34635 i $ & $-0.757591 + 7.39531 i$ & $ -9.83135 + 17.1882 i$ \\ \hline

3 & $ 0.572121 + 5.37043 i $ & $-4.47751 + 0.239986 i$ & $ 129.784 - 108.923 i$ & $ -120.673 + 140.163 i$ \\ \hline

\end{tabular}

\caption{Scattering coefficients for $\sigma \leftrightarrow -\sigma $ scenario where $m'=1$.} \label{sftab2}

\end{table}

\begin{figure}[H]
  \centering
   \includegraphics[width=0.45\textwidth]{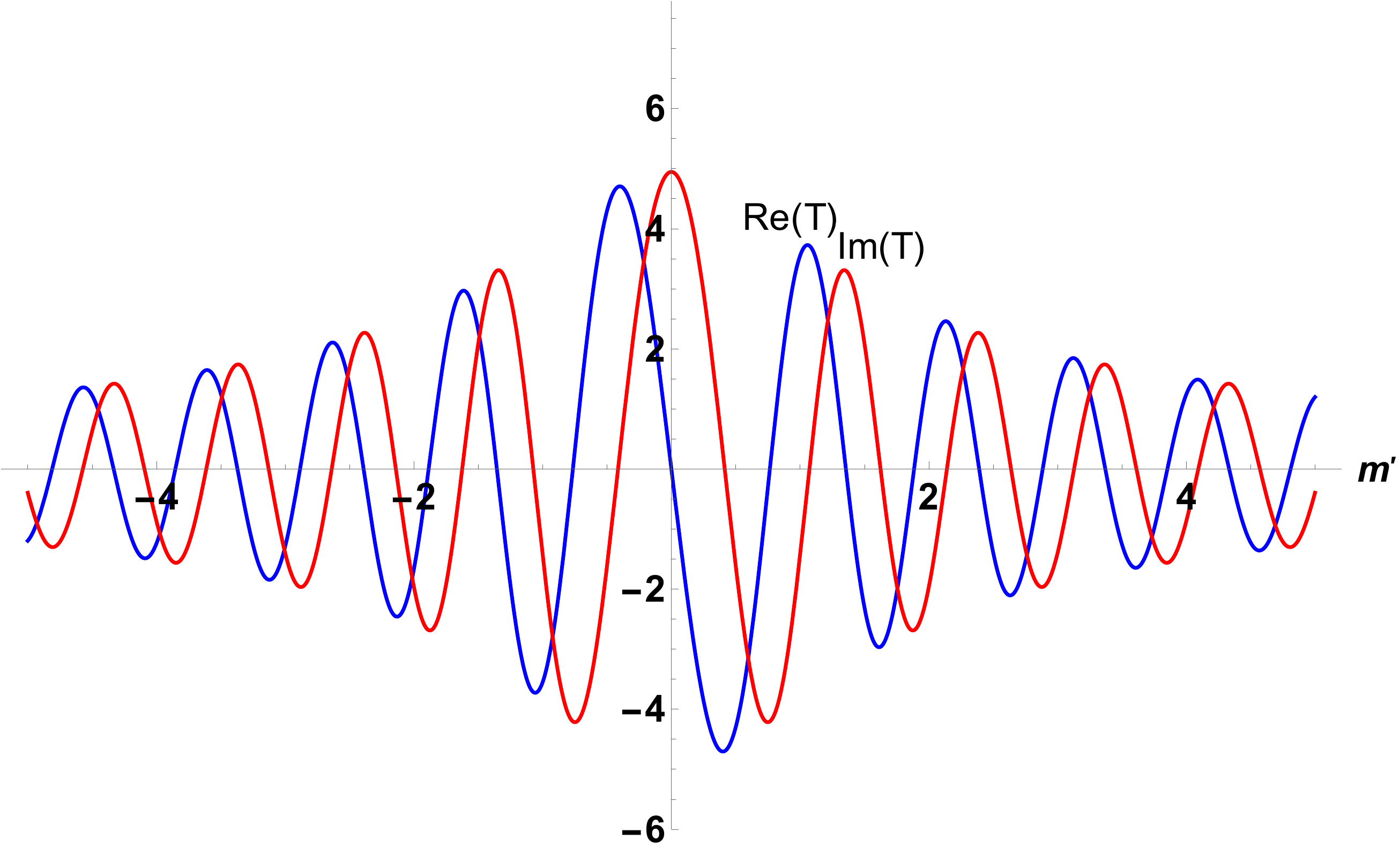}
   \includegraphics[width=0.45\textwidth]{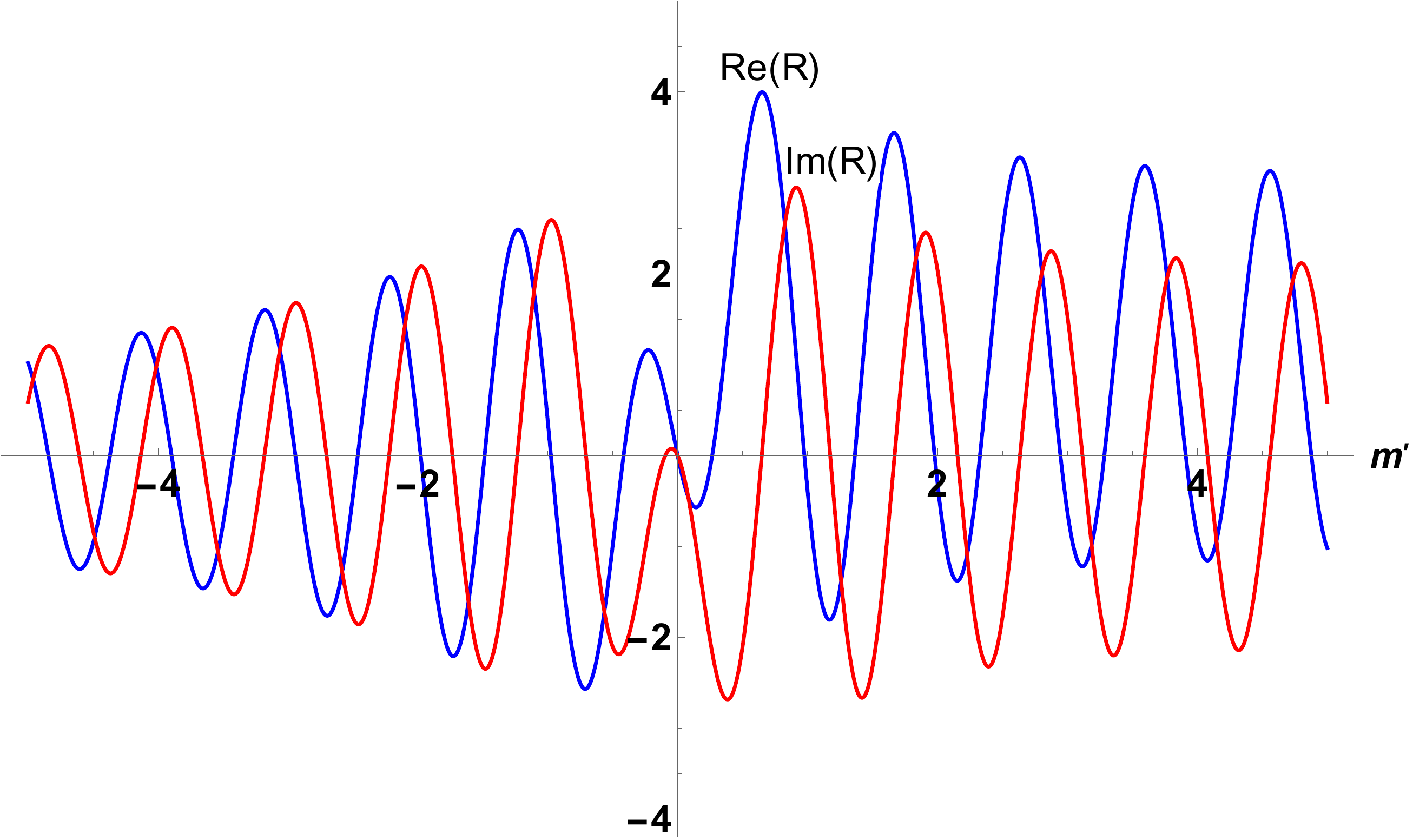}
  \caption{The $m'$-dependence of the transmission coefficient $T$ and reflection coefficient $R$ in the scattering with spin-flip scenario.}
\label{TRmSFfig}
\end{figure}

Another possibility for scattering that results in the outgoing waves in the lower plane is to replace $H^{(1)}(kR(u))$ with $H^{(2)}(-kR(u))$ in the lower plane region of Eqn.~(\ref{eq:BC01}) and (\ref{eq:BC02}). All of the scattering coefficients numerically solved turn out to be the same except the change in the transmission coefficients $T\to -iT$ for each corresponding case.  The wave function profile in the three regions are also identical in every case.

\subsection{Comments on fermionic QNMs of wormhole}  \label{secQNM}

Table~\ref{Tabel:ConFlux} shows QNMs/energies of the fermion in the magnetized wormhole for all combinations $(\kappa_{1},\kappa_{2})=(+1,+1),(-1,-1),(+1,-1),(-1,+1)$. We can see from the expression that the imaginary and real parts of energy depend on $n,m',M$ as well as the spin $\sigma$. For $(\kappa_{1},\kappa_{2})=(+1,+1),(-1,-1)$ regardless of the spin, $E$ is real for $n+1, n \leq \displaystyle{\frac{Mcr}{\hbar}}$ and pure imaginary otherwise. Note the reality of $E$ depends on the ratio of the radius of curvature in $u$ direction, $r$, and the Compton wavelength $\hbar/Mc$ of the fermion.

Figure~\ref{QNMfig} shows dependency of $E$ on each parameter when the spin is chosen to $\sigma =1$ for $(\kappa_{1},\kappa_{2})=(+1,-1),(-1,+1)$. It has certain similar characteristics to the QNMs investigated in Ref.~\cite{Lasenby:2002mc,Cho:2003qe,Jing:2005dt,Blazquez-Salcedo:2017bld,Blazquez-Salcedo:2018bqy}. Namely for some range of parameters, $\text{Im} (E)$ could be very small. These long-lived fermionic QNMs are quite common for both black holes and wormholes. For our wormhole, the long-lived modes are the ones close to the Re$(E)$ axis in each plot of Fig.~\ref{QNMfig}. While Re$(E)$ decreases with $n$ and increases with $m'$\&$M$, $|\text{Im}(E)|$ increases with $n\&m'$ and decreases with $M$.

\begin{figure}[H]
  \centering
   \includegraphics[width=0.4\textwidth]{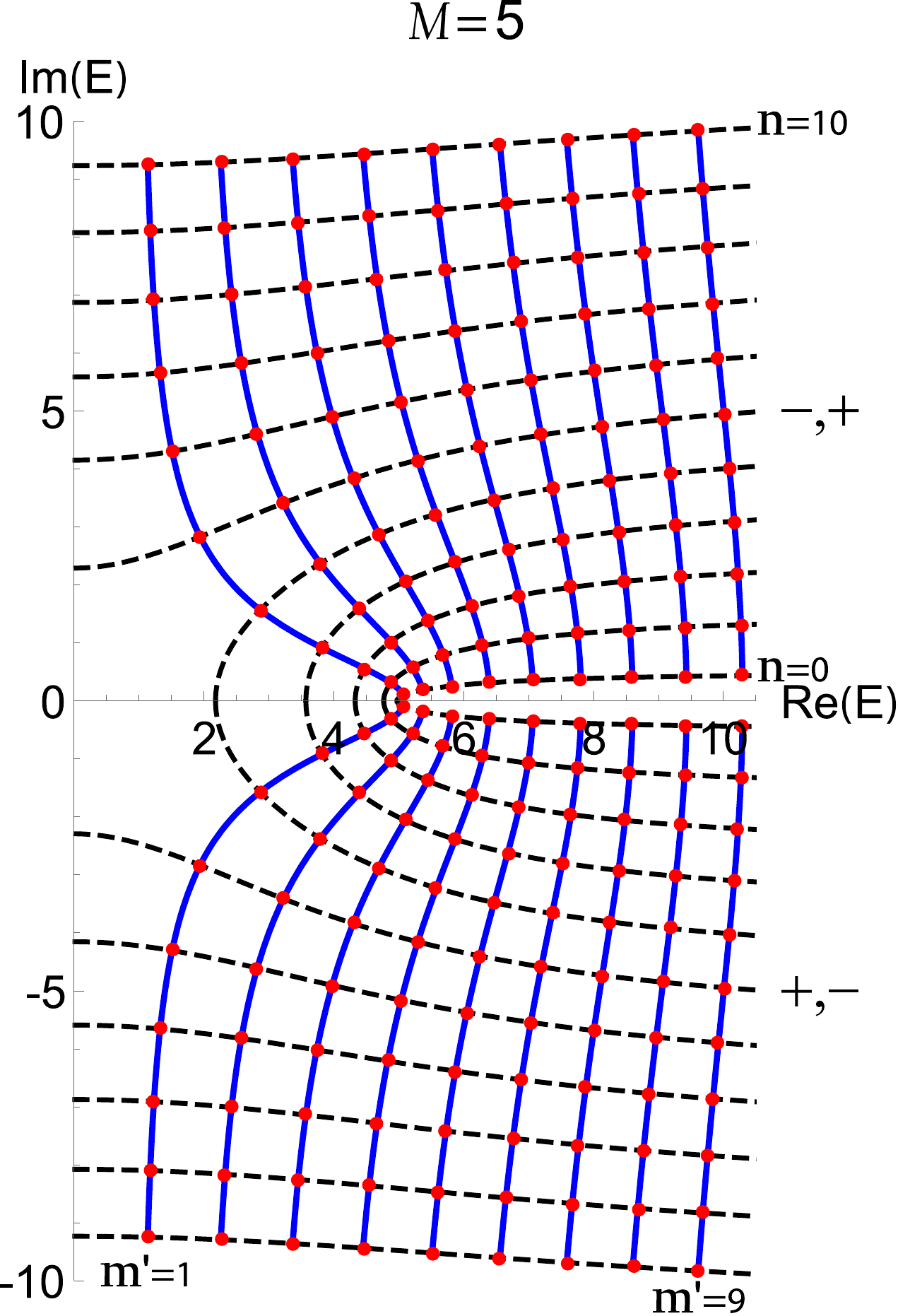}
   \includegraphics[width=0.49\textwidth]{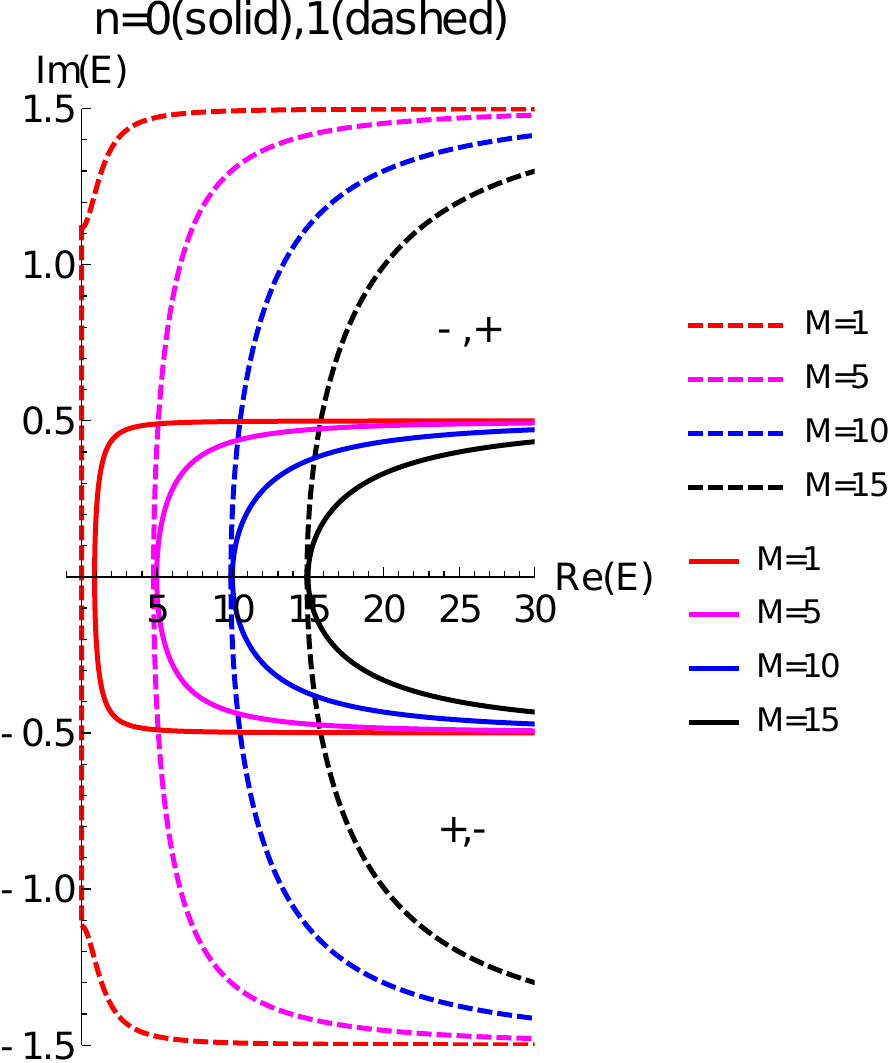} \\
   \includegraphics[width=0.4\textwidth]{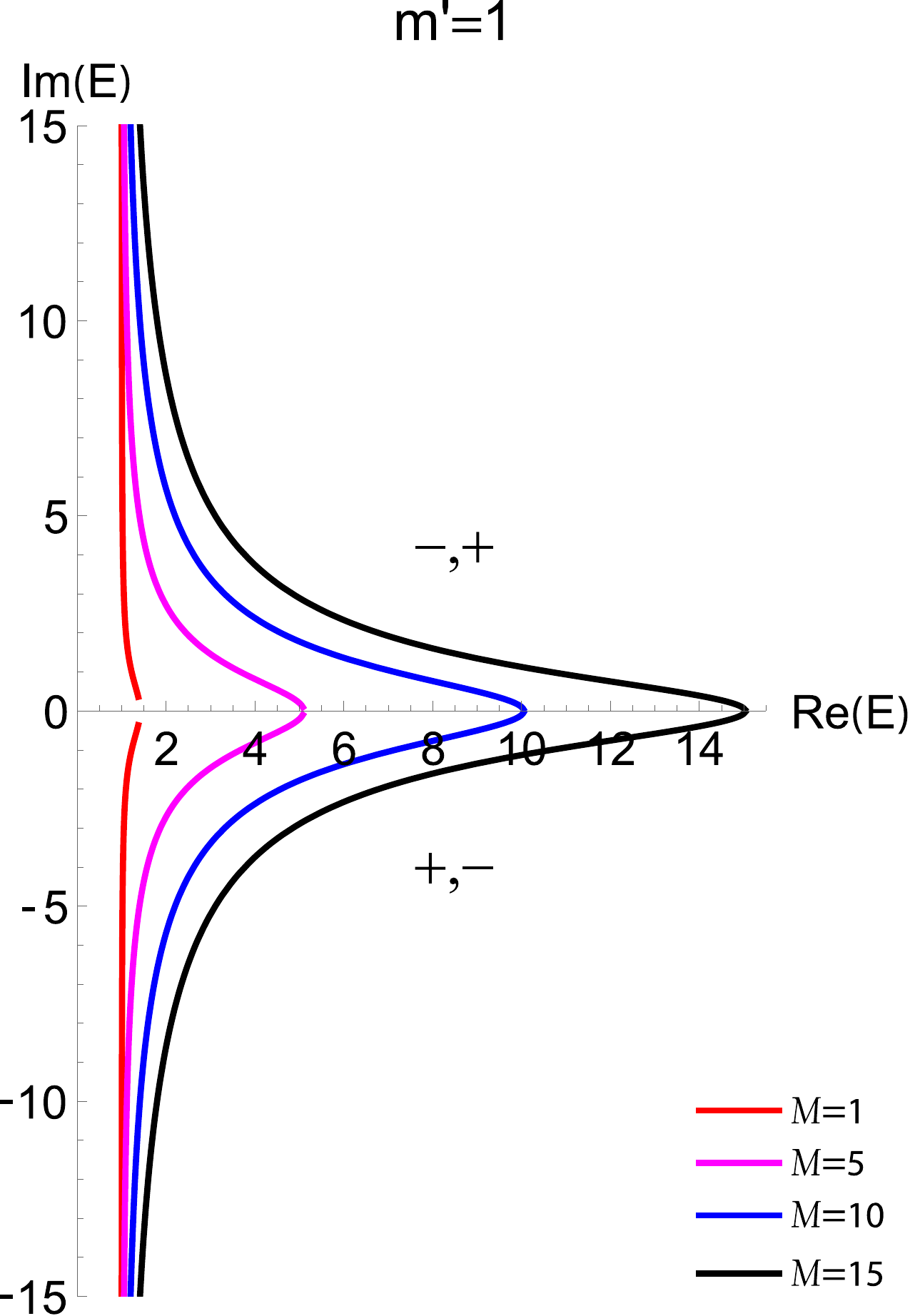}
   \includegraphics[width=0.4\textwidth]{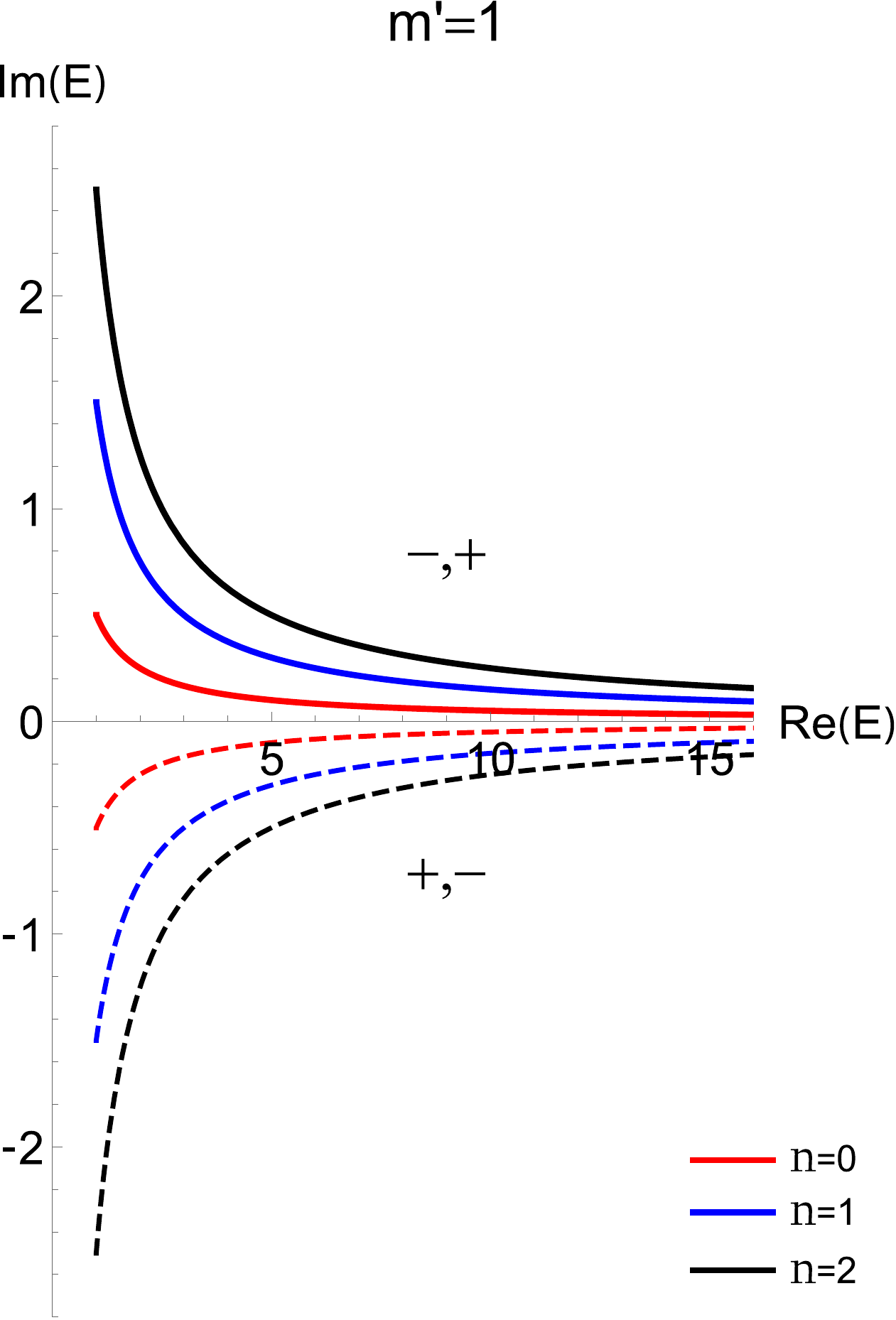}
  \caption{The QNMs $E=\hbar\omega$ of fermion on the wormhole from Table~\ref{Tabel:ConFlux} in natural units for $(\kappa_{1},\kappa_{2})=(+1,-1),(-1,+1)$.}
\label{QNMfig}
\end{figure}

In Ref.~\cite{Blazquez-Salcedo:2019uqq}, massless and massive wormhole supported by fermionic fields are constructed with diverging field profile~(divergent total fermion density integral). Interestingly, the massless wormhole has no time dilation similar to the wormhole we consider in this work. However, in our case the wormhole is assumed as background presumably constructed from other materials where the back reaction to the surface is neglected. Certain classes of our fermionic solution~(the ones with QNMs) actually diverge in the wormhole space if it is continued to infinity. The back reaction should be large but results of Ref.~\cite{Blazquez-Salcedo:2019uqq} suggests that this divergence might actually be consistent with different wormhole geometry. This interesting aspect will be explored in the future work. 

\section{Discussions and Conclusions}  \label{secCon}

The scattering of fermion to the 2-dimensional wormhole is calculated by considering both normal and quasinormal modes of the fermionic states in the wormhole.  For incoming waves with real momentum and energy, we found the quantum selection of the states allowed to propagate through the wormhole, i.e., the momentum-spin relation given by Eqn.~(\ref{kmrel}). Only the states with the quantized momentum related to its angular momentum can be transmitted through the hole via the normal modes and thus be maintained stationary for a long period of time. The waves will also be partially reflected with the same momentum and energy. The unitarity condition (\ref{urel}) is valid in this case. Notably, the main result, Eqn.~(\ref{kmrel}), is also valid for uncharged or zero-flux fermion scattering with a replacement $m'\to m$.

For scattering involving QNMs of the wormhole, the process will be decaying or growing in time due to the imaginary part of the frequency. For QNMs with zero imaginary part, e.g. when $m'=0$~(this is not exactly a normal mode since the momentum is not real), the energy can be real for massive fermion and sufficiently small $n$.  Even in such cases, the momentum will be pure imaginary resulting in spatial attenuation and enhancement of the wave functions in the upper and lower plane regions. The scattering with QNMs of the wormhole violates unitarity but satisfies more generic relation (\ref{TRrel}). Only the incoming waves with the right complex energy and momentum $E=E_{n,m'}, k=k_{n,m'}$ will undergo a resonant scattering with the hole resulting in partially transmitted and reflected waves. The physical energy of the particle, however, is given by the real part of $E$. Remarkably, there will be only transmitted (tunneling) waves with no reflection for $m'=0$ fermion scattering, i.e., when the magnetic flux is quantized in integer multiples of the magnetic flux quantum $\phi= mhc/e$. For general nonzero $m'$ scenario, the transmission and reflection coefficients have curious peaks at $m'=\pm 6$

Lastly following the argument in Ref.~\cite{Rojjanason:2018icy}, a few comments on fermion scattering in the graphene wormhole can be made. With $c\to v_{F}\simeq 10^{6}{\rm m/s}$ and $M=0$ replacement, the normal-modes energy of the fermion in the graphene wormhole is $\displaystyle{E=\frac{\hbar v_{F}m'}{a\sqrt{q}}}=\displaystyle{\frac{0.658~{\rm nm}}{a}\frac{m'}{\sqrt{q}}}$ eV. Only quasi-fermion with this quantized energy can be transmitted through the graphene wormhole as a stationary state, and it will be accompanied by the reflected waves obeying the unitarity condition. The effects of stitching the graphene wormhole to the plane include the exchange of inequivalent Dirac points that can be taken into account by the effective flux of gauge field~\cite{Gonzalez,Garcia:2019gro}. This effective flux can simply be added to the magnetic flux in our work resulting in the change of the effective angular momentum number $\displaystyle{m'=m-\frac{\phi_{\rm total}}{\phi_{0}}}$ and otherwise the same results.

\section*{Acknowledgements}

P.B. is supported in part by the Thailand Research Fund (TRF),
Office of Higher Education Commission (OHEC) and Chulalongkorn University under grant RSA6180002.

\appendix

\section{Wave function profiles}

\begin{figure}[ht]
  \centering
  \subfigure[real $k$]{%
\includegraphics[width=0.45\textwidth]{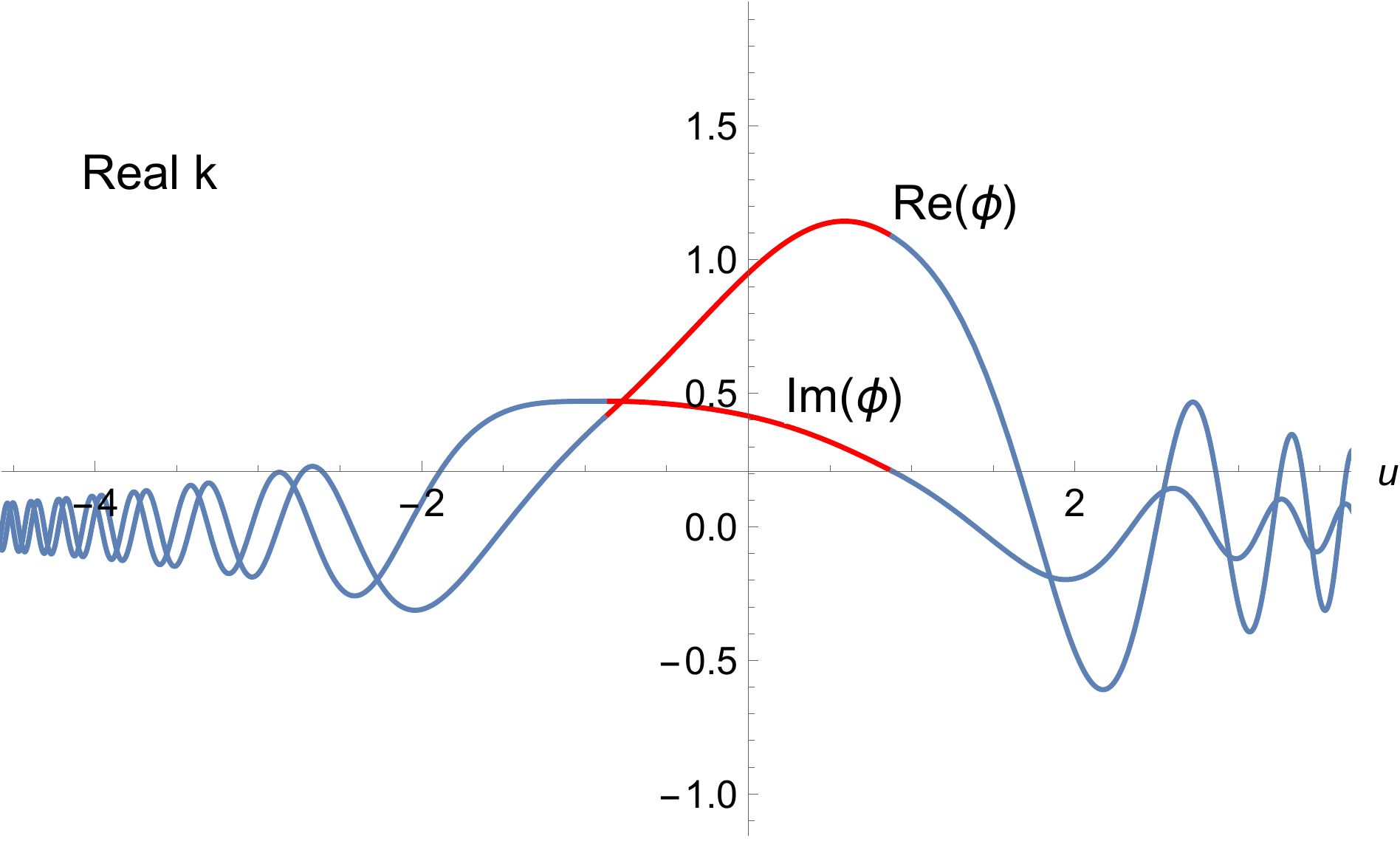}
\label{realkfig}}
\quad
\subfigure[$m'=0$]{%
\includegraphics[width=0.45\textwidth]{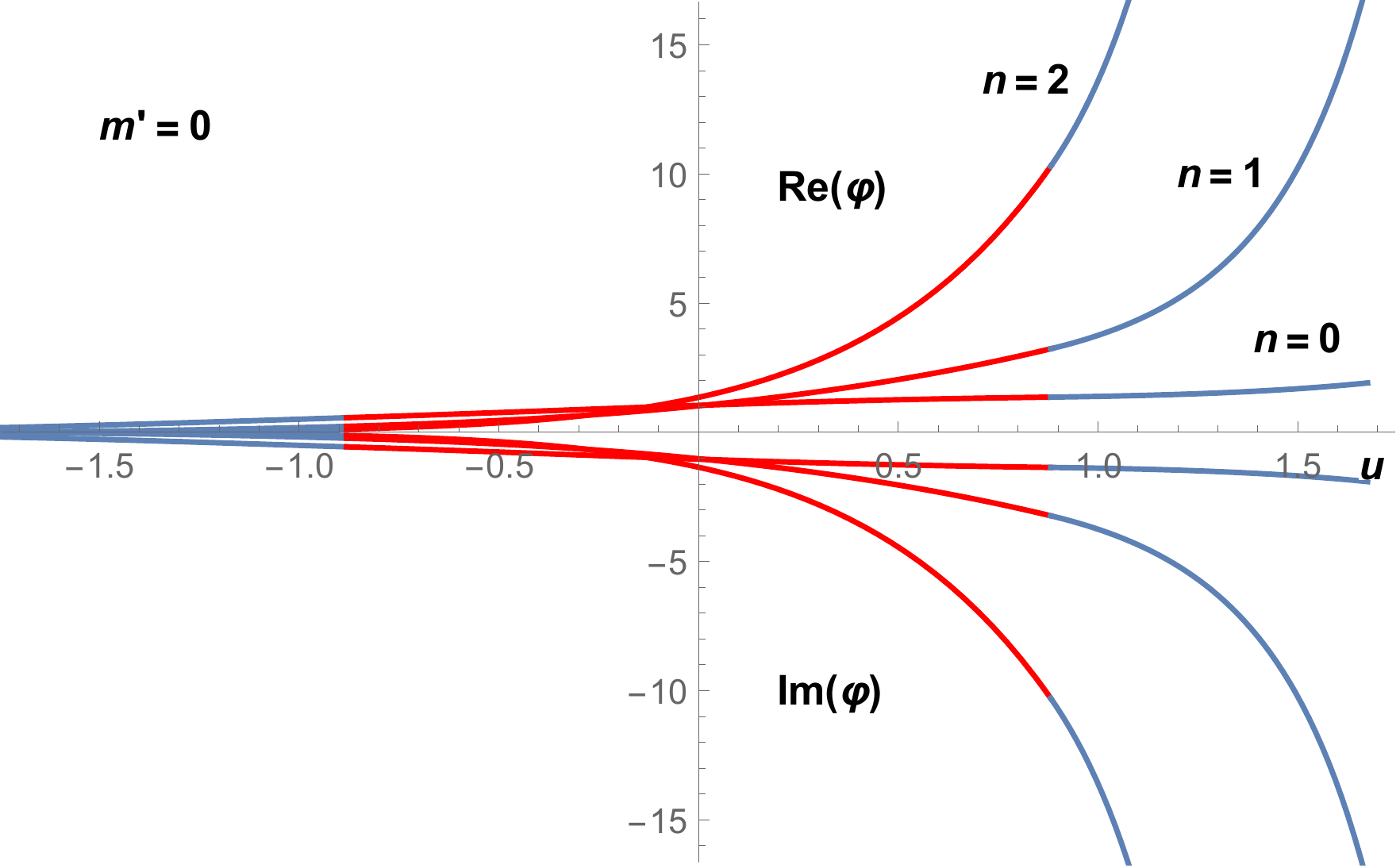}
\label{m0fig}}   
\subfigure[general $m'$]{%
\includegraphics[width=0.45\textwidth]{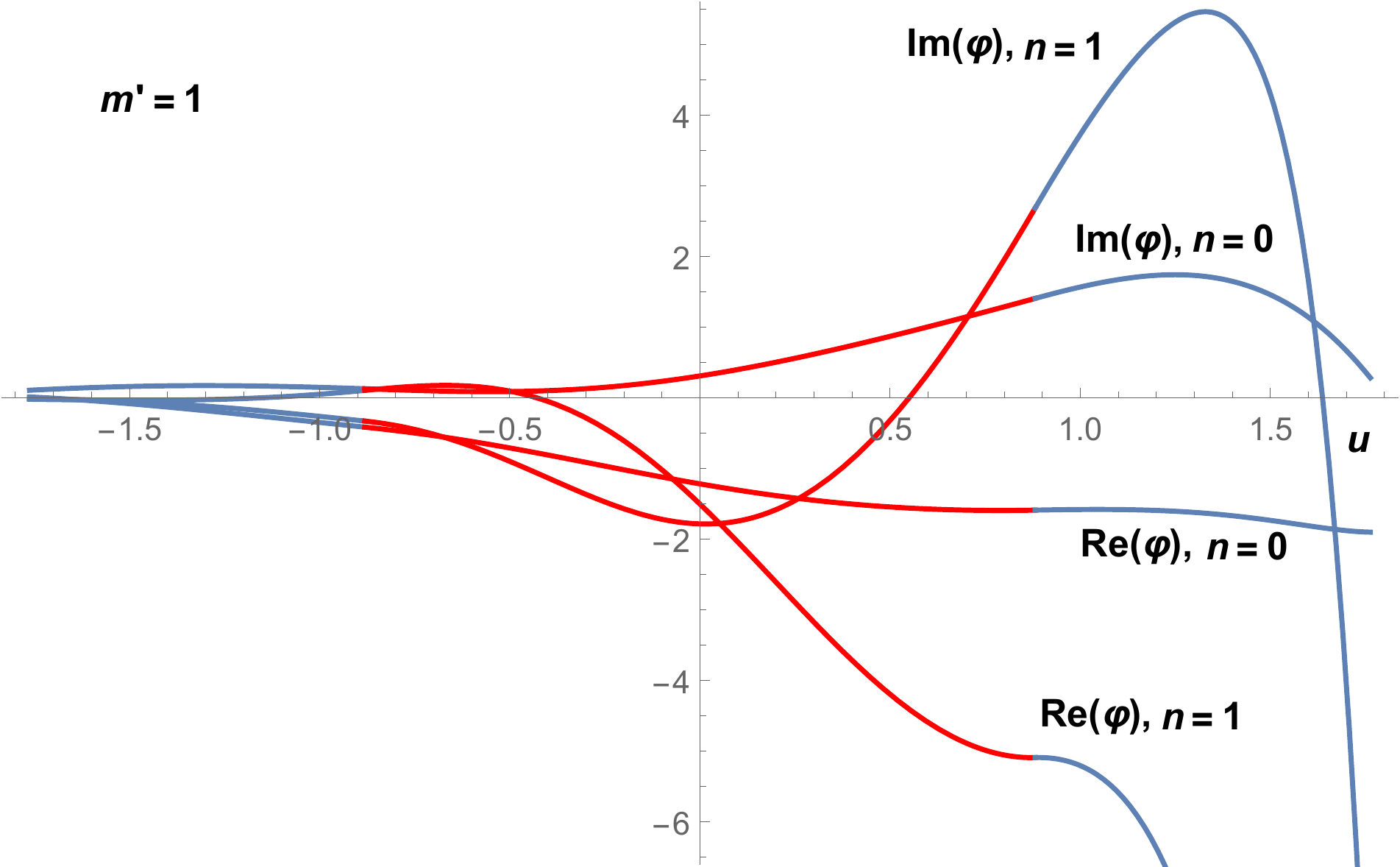}
\label{genmfig}}
\quad
\subfigure[spin flip]{%
\includegraphics[width=0.45\textwidth]{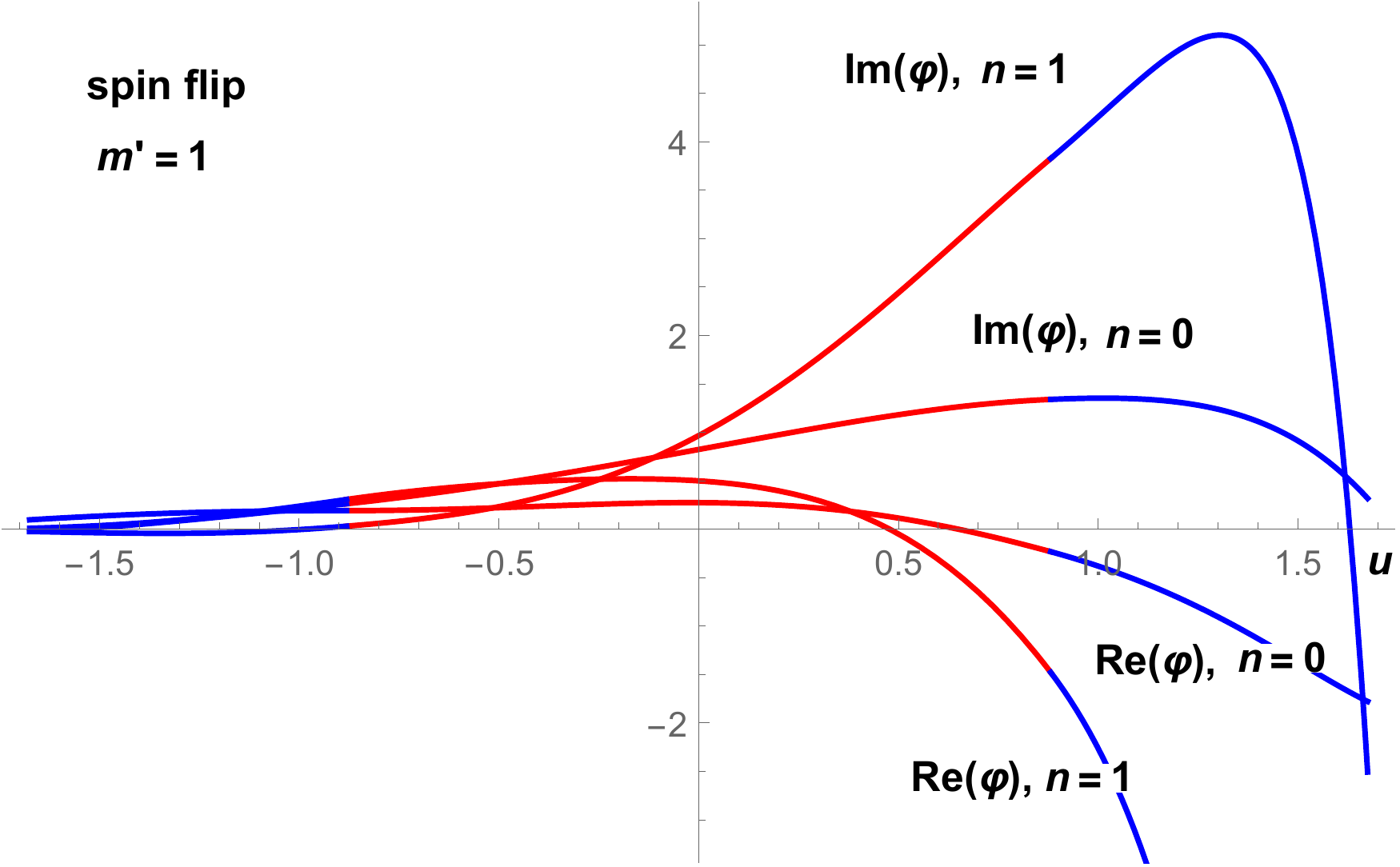}
\label{Spinflipfig}}

  \caption{Wave function profiles of scattering fermion for (a) real momentum, (b) $m'=0$, (c) general $m'$, and (d) spin-flip scenario. The real and imaginary parts of the wave functions in the wormhole region $u_{m}<u<u_{p}$ are depicted in red. }
\end{figure}


\end{document}